\documentclass[10pt]{article}

\usepackage{graphics}
\usepackage{epsfig}
\usepackage[]{amsfonts,amsmath,amsthm,amssymb,amscd}
\newcommand{\bea}{\begin{eqnarray}}

\newcommand{\bsa}{\begin{subeqnarray}}
\newcommand{\esa}{\end{subeqnarray}}

\newcommand{\rf}[1]{(\ref{#1})}

\newcommand{\bF}{\mathbf{F}}
\newcommand{\bq}{\mathbf{q}}
\newcommand{\bu}{\mathbf{u}}

\newcommand{\br}{\mathbf{r}}

\newcommand{\be}{\mathbf{e}}

\newcommand{\ez}{{\varepsilon}}

\newcommand{\hu}{\widehat{u}}
\newcommand{\hv}{\widehat{v}}

\newcommand{\ds}{\displaystyle}

\newcommand{\xtext}[1]{\mbox{#1}}

\newcommand{\integer}[1]{\left\lfloor{#1}\right\rfloor}

\newcommand{\ssx}{d}
\newcommand{\ssy}{d}

\begin{document}
\title{ {\bf Kinetic Monte Carlo Simulation of Strained Heteroepitaxial Growth with Intermixing}}


\author{Arvind Baskaran \\ 
              Department of Mathematics, University of Michigan, Ann Arbor, MI, USA \\
              baskaran@umich.edu           \\
        Jason Devita  \\
              Department of Mathematics, University of California Los Angeles, CA, USA\\
              jason@math.ucla.edu         \\
        Peter Smereka\\
              Department of Mathematics, University of Michigan, Ann Arbor, MI, USA \\
              psmereka@umich.edu        \\
}



\maketitle

\begin{abstract}
An efficient method for the simulation of strained heteroepitaxial growth 
with intermixing using kinetic Monte Carlo is presented. The model used is based on a 
solid-on-solid bond counting formulation in which elastic effects are incorporated
using a ball and spring model. While idealized, this model nevertheless captures 
many aspects of heteroepitaxial growth, including nucleation, surface diffusion,
and long range effects due elastic interaction. The algorithm combines a 
fast evaluation of
the elastic displacement field with an efficient implementation of a rejection-reduced 
kinetic Monte Carlo based on using upper bounds for the rates.
The former is achieved by using a multigrid method for global updates of the
displacement field and an expanding box method for local updates.
The simulations show the importance of intermixing on the growth of a strained
film.  Further the method is used to simulate the growth of self-assembled stacked quantum dots.
\end{abstract}

\section{Introduction}
Heteroepitaxy is the process of slow deposition of a film of one or more crystalline
materials on a crystalline substrate of a different material. 
The classical examples of film/substrate combinations include, for example
Ge/Si, InAs/GaAs, and InP/GaAs. The natural lattice spacing between the
film and the substrate differ by a few percent resulting in elastic stress.
Consequently, as the film grows the elastic energy builds up. 
The formation of 3D islands reduces the elastic energy by relieving stress 
at the cost of increased surface energy. In many cases these
islands are on the order of tens of nanometers and are referred to as 
quantum dots.  These quantum dot materials are of importance in 
construction of some optoelectronic devices.
 
The theory of island formation is well understood in the context of 
Asaro-Tiller-Grinfeld instability \cite{asaro_tiller,grinfeld}. 
This explains the instability of a single component stressed film to
perturbations of the flat surface.  The flat surface is unstable 
under sufficiently long wavelength perturbations. 
These perturbations grow through mass transport on the surface, 
along the free energy gradients, leading to formation of 
surface ripples that later grow into large islands
with stress concentration in the valleys.  The theory however fails to 
explain an experimental observation, namely in many cases the
film first grows in a layer-by-layer fashion until
it reaches a certain critical thickness. 
Then islands form on top of this layer (known as the wetting layer). 
This mode of growth is known as the Stranski Krastonov (SK) 
growth mode\cite{SK_1,SK_2}.
 
It was demonstrated experimentally by Cullis and co workers 
\cite{cullis_InGaAs_exp,cullis_InGaAs_sim},  
that the intermixing  between the film and substrate is of importance. 
The presence of compositional non-uniformity and the dilution of the
film by substrate material can have a stabilizing effects. 
Moreover the presence of vertical and lateral segregation in 
the case of alloy films \cite{cullis_lateral_vertical_exp}, 
can further change the nature of the instability. 
Subsequently, Tu and Tersoff\cite{tersoff_tu}, using a continuum
model developed by Spencer et al\cite{spencer_vorhees_tersoff},
argued that Stranski Krastonov growth is a kinetic effect and
the wetting layer is not stable. A crucial aspect of their
proposal is the intermixing that occurs between the film
and the substrate.

Another important example in which intermixing occurs on
very large length scales is the self assembly of stacked quantum dots. 
Here the dominant mechanism of intermixing results from
the deposition process in which
several layers of one material deposited which are followed
by several layers of another material and so on. Since the materials
are not lattice matched elastic strain develops and it is observed that
the film forms quantum dots. 
The quantum dots in different layers align themselves 
to self assemble into stacked quantum dots. 
Intermixing of the film material
and the capping layer needs to be addressed in this situation.
Some related work, both experimental and computational,
can be found in
Refs. \cite{tersoff_tu_new,JMM,gold,Ratsch,voorhees_thornton,quek_lui_phasefield}. 

It is clear that the simulation of strained heteroepitaxial growth with
intermixing has important applications. One reasonably popular 
approach is based on numerical solution of continuum equations.
Indeed recent work in this direction
\cite{spencer_vorhees_tersoff,tersoff_tu,tersoff_tu_new,wise_lowengrub_kim,voorhees_thornton} 
has been able to capture many aspects of the film growth. Another
approach is based on kinetic Monte Carlo (KMC). Since KMC is
based on an atomistic scale it is typically slower than a continuum
formulation. On the other hand, KMC not only captures all the
physical effects modeled by continuum approach but can naturally
include discrete and stochastic effects such as nucleation,
surface roughness, and intermixing.

The purpose of this paper is to present a KMC model
of strained heteroepitaxial growth with intermixing and
an algorithm for its efficient simulation. We extended the model proposed by
Orr et al \cite{orr} and Lam, Lee, \& Sander \cite{lam_lee_sander} to
include intermixing. This model is a solid-on-solid bond counting scheme\cite{VV} in which
elastic interaction are included using a ball and spring model. 
The efficient simulation of this model represents the major contribution of
this paper.

One computational bottle neck for the simulation of strained film growth is the
repeated calculation of elastic field. In this paper we  address this
by  the inclusion of intermixing into the multigrid-Fourier method developed by 
Russo \& Smereka\cite{russo_smereka_3D,russo_smereka_2D}. In addition
we present a cleaner way to incorporate the substrate into the
multigrid method and simplify the coarse-graining
and prolongation operators.  As it turns out,  for the KMC model used here, the
change in total elastic energy, when a surface
atom is removed, is needed. It was established by Schulze \& Smereka\cite{schulze_smereka}
that despite the long range nature of elastic interactions one can accurately compute
this energy change by local calculations. In this work, we have verified that 
this approach also works in case of intermixing. In addition we have presented
numerical evidence that the asymptotic expressions presented in
Ref. \cite{schulze_smereka} remain valid in the case of intermixing.

Another computational bottleneck is the computation of the rates.
In order to implement rejection-free KMC one must know the hopping
rates for all the atoms. To accomplish this one would have to
compute the change in elastic energy  that occurs when 
each and every atom is removed. This would be prohibitively expensive.
Fortunately, it was argued  in Ref. \cite{schulze_smereka} that
the  current elastic energy density can be used to provide fairly sharp upper
bounds on the change in elastic energy. These can be used to provide
a rejection-reduced KMC algorithm. Here we demonstrate that this
approach works well for the intermixing case.  In our simulations 
the rejection rate is approximately between .5 and 5 percent.

In the following sections we explain the KMC model 
and present our algorithm to calculate the elastic displacement field 
and  elastic energy in the  case of a multicomponent film.
Some results are presented that illustrate the effectiveness of the algorithm; 
namely,  the effects of intermixing on the morphology of a growing strained film
and a simulation illustrating the self-assembly of 
stacked quantum dots. To the best of our knowledge these results are
the first kinetic Monte Carlo simulations of strained heteroepitaxial growth with intermixing.

\section{Model Description}
Our model is based on the solid-on-solid model presented
by  Orr et al \cite{orr} and Lam et al \cite{lam_lee_sander}.
We consider a semi-infinite substrate
which is initially  composed of a single component. For ease of exposition
we shall refer to the substrate material as Silicon. The deposited material
will be composed of a prescribed mixture of Silicon and Germanium.
The model and algorithm will be detailed for a two component film but 
it is not limited to this.  In this paper we restrict ourselves to 1+1 dimensions. 
The atoms in the crystal occupy sites on a simple cubic crystal,
within the solid-on-solid framework (no overhanging atoms). Therefore
the height of the surface is a function of the horizontal coordinate
denoted by $\ell$.
Each atom in the lattice is bonded with its nearest (4 possible) and
next to nearest neighbors (4 possible) by bonds of strength $\gamma$. 

The elastic interactions are modeled by means of a ball and spring system. 
Each atom on the lattice is connected to its nearest and next to nearest 
neighbors by Hookean springs.  The lateral and vertical springs have 
a spring constant of $k_L$ and the diagonal springs $k_D$. 
We choose $k_D= k_L/2$ corresponding to the isotropic case 
\cite{lam_lee_sander}. 
The natural bond lengths (lattice spacing) are denoted by $a_{ss}$,  $a_{sg}$ 
and $a_{gg}$, for Si-Si, Si-Ge and Ge-Si bonds respectively. 
If these quantities are equal there are no forces due to lattice mismatch 
and hence no elastic energy in the crystal. 
In general since $a_{ss} \neq a_{sg} \neq a_{gg}$ forces do arise.
This is addressed in detail in the next section.

The crystal evolves by rearranging itself through motion of surface atoms.
By virtue of the solid-on-solid assumption each surface atom is uniquely
specified by $\ell$ ( the horizontal component) and this is
what we mean when we specify the $\ell$th surface atom.
The hopping rate of the $\ell$th surface atom is modeled by 
\begin{equation}
\label{rate_eqn}
R_\ell=R_0 \exp \left[\frac{\Delta E + E_0}{k_BT}\right]
\end{equation} 
where
$\Delta E$ is the change in the total energy when removing the $\ell$th surface atom, 
$R_0$ is the attempt frequency, $k_B$ is the Boltzmann constant and 
$T$ is the absolute temperature. We write the total energy as
\[
E = E_{chem} + W
\]
where $E_{chem}$ is the total chemical energy and $W$ is the total elastic
energy. The chemical energy can be thought of as the contribution to
the total energy from local interactions whereas $W$ is the contribution
by the long range interactions. Since we are assuming that the bond
energy is the same for nearest and next to nearest neighbors, it follows that
\[
\Delta E_{chem}=-\gamma N
\]
where $N$ is the total number of nearest and next to nearest neighbors.
Therefore, the change in total energy is given by
\begin{equation}
\Delta E =-\gamma N + \Delta W
\end{equation}
The calculation of $\Delta W$ is described in the next two sections. 
The parameters $E_0$ and $R_0$ will be chosen to match the adatom diffusion
rate to experimental values. When an atom hops it moves to another 
surface site whose horizontal coordinate changes by $\pm 1$,
with equal probability. As a simplification we
shall ignore the elastic contribution to the energy barrier for
adatoms; therefore the rates we shall use are
\begin{equation}
R_\ell =\left\{ 
\begin{array}{ll}
 R_0 \exp \left({\displaystyle \frac{-3 \gamma + E_0}{k_B T}} 
\right) &\quad\xtext{if}\quad  N\leq 3 
\\[.2in]
R_0 \exp \left(\displaystyle{ \frac{-N \gamma  + \Delta W + E_0}{k_B T}} 
\right) &\quad \xtext{if}\quad N>3 
\end{array} \right.
\label{rate_normal_eqn}
\end{equation}
The number 3 corresponds to the number of bonds of an adatom on a flat terrace.
The adatoms with less than 3 bonds are assumed to hop at the same rate as other adatoms. 
This will reduces the computational time significantly. 
No significant change is observed in the results due to this treatement for the adatoms.  

Finally we deposit atoms at a rate of 
\begin{equation}
\label{rate_del}
R_{dep} = F M 
\end{equation}
where $F$ is the deposition rate in monolayers per unit time
and $M$ is the total number of sites in the horizontal direction.

The model by Lam et al \cite{lam_lee_sander} allowed the possibility of
large hops with $R_0$ adjusted to correct for 
the large hop size.
This greatly improved the computational speed
but at the same time not changing the results too much.
However in the case of intermixing we observed that the
inclusion of this feature significantly retarded intermixing. 
Hence large hop strategy was not adopted in this work.

\section{The Displacement Field}
The calculation of the hopping rate of an atom involves the computation of 
elastic energy of the crystal in different configurations. 
The elastic energy is merely the sum total of the energy stored in each 
spring when the crystal is in mechanical equilibrium. 
The energy in a harmonic spring is proportional to the square of the 
change in length of the spring (relative to natural length). 
This is computed in terms of the displacements of the atoms from 
a reference configuration. The computational challenge is quickly
computing the equilibrium displacements of the atoms.

\subsection{Forces In the Reference Configuration}
The first step toward the calculation of the elastic energy is the choice of a 
reference configuration. 
The displacements of atoms are calculated with respect to this reference configuration.
We choose this to be a cubic lattice, whose lattice spacing equals that of pure 
substrate material. 
This ensures that the substrate is devoid of any forces in the reference state. 
Thus two atoms in the reference configuration are a distance $a_{ss}$ apart. 
The natural lattice spacing however depends on the nature of the two atoms. 
The natural lattice spacing is $a = a_{ss}, a_{gs}$, and $a_{gg}$, when the 
two atoms are Si-Si, Ge-Si, and Ge-Ge respectively.  It is useful
to introduce the misfit parameter, $\epsilon$, 
\begin{equation}
\label{misfit}
\epsilon = \frac{(a - a_{ss})}{a_{ss}}
\end{equation}
This typically varies from $\epsilon = -0.06 \mbox{ to }+0.06$. 
In the two component system Ge/Si we have two interactions to be considered Ge-Si and Ge-Ge 
and we introduce the following quantities
\begin{equation}
\label{misfit_values}
\epsilon_{sg} = \frac{(a_{sg} - a_{ss})}{a_{ss}} \quad \mbox{and}\quad 
\epsilon_{gg} = \frac{(a_{gg} - a_{ss})}{a_{ss}}.
\end{equation}

To compute the forces, that arise from the misfit, on the atoms in the reference configuration
let us focus on two nearest neighbor atoms.
The force exerted by the spring on the atoms can be calculated to 
be $k_L ( a - a_{ss} )$ in magnitude where $k_L$ is the spring constant
for nearest neighbor atoms.
In the case of next to nearest neighbor interaction, the force is
calculated to be $k_D ( a - a_{ss} )$ in magnitude,  to first order in $\epsilon$.
The spring constant for next to nearest neighbors is $k_D$. Before
the general formula for the forces in the reference configuration are
given, we first present an simple example.
 
\begin{figure}[!ht] 
\begin{center}
\begin{picture}(200,180)
\includegraphics[width=2.2in]{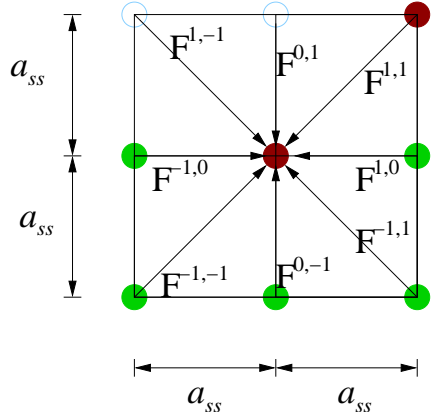}
\end{picture}
\end{center}
\caption{Configuration shows the atom of interest in the center surrounded by 8 neighboring sites. The Green circles represent Si atoms, the Red represent Ge and Blue the empty sites.}
\label{fig_config_demo_forces}
\end{figure}

\medskip
\noindent
{\it An Example.}
In order to make things clear we present a sample calculation of forces 
in the configuration shown in Fig: \ref{fig_config_demo_forces}.
The forces are given by
\begin{equation}
\bF^{-1,1}=\left(  \begin{array}{c} 0 \\[.05in] 0 \end{array}\right), \quad
\bF^{0,1}=\left(  \begin{array}{c} 0 \\ [.05in]0 \end{array}\right), \quad\mbox{and}\quad
\bF^{1,1}=\left(  \begin{array}{c} -F^{gg}_{NN} \\[.05in] -F^{gg}_{NN} \end{array}\right).
 \end{equation}
The first two correspond to interactions with empty sites and the last one arises
from a Ge-Ge interaction.  The forces from Si-Ge interactions are given by
\begin{equation}
\begin{array}{ccc}
\bF^{-1,0}=\left(  \begin{array}{c} F^{sg}_N \\[.05in] 0 \end{array}\right),\quad & 
\phantom{\bF^{-1,0}=\left(  \begin{array}{c} F^{sg}_N \\[.05in] 0 \end{array}\right),\quad} & 
\bF^{1,0}=\left(  \begin{array}{c} -F^{sg}_N \\[.05in] 0 \end{array}\right), \\[.2in]
\bF^{0,-1}=\left(  \begin{array}{c} 0 \\[.05in] F^{sg}_N \end{array}\right),\quad &
\bF^{-1,-1}=\left(  \begin{array}{c} F^{sg}_{NN} \\[.05in] F^{sg}_{NN}  \end{array}\right),\quad &
\bF^{1,-1}=\left(  \begin{array}{c} F^{sg}_{NN} \\[.05in] -F^{sg}_{NN} \end{array}\right). 
\end{array}
\end{equation}
where
\begin{equation}
F^{sg}_N = k_L \epsilon_{sg} a_{ss},  \quad F^{sg}_{NN} = k_D \epsilon_{sg} a_{ss}, \quad \mbox{and} \quad  F^{gg}_{NN} = k_D \epsilon_{gg} a_{ss}. 
\end{equation}

\medskip
\noindent
{\it The General  Case.}
The crystal is a lattice indexed by $(\ell,j)$. 
The first index represents the horizontal coordinate. 
The crystal is periodic in this index. 
The second is the vertical coordinate. 
The crystal is semi-infinite in this index, 
i.e, $j<h_{\ell}$ where $h_{\ell}$ is the location of the surface. 
All sites $ j>h_{\ell}$ are unoccupied.

We now write down the misfit forces for the general case.
The net force on an atom at site $(\ell,j)$ in the reference state, due to its nearest and next to nearest neighbors $ ( j+m, \ell+n ) $ where $ (m,n) \in (\{ -1, 0, 1 \}, \{ -1, 0, 1 \} )$, is given by
\begin{equation}
\label{netforce}
\left( \begin{array}{c} F_{\ell j} \\ G_{\ell j} \end{array} \right) = 
\sum_{m,n=-1,0,1} \bF_{\ell j}^{mn}
\end{equation}
where $\bF_{\ell j}^{mn} $ are given by 
\begin{equation}
\begin{array}{ccc}
\bF_{\ell j}^{-1,1}=\left(  \begin{array}{c} f^{-1,1}_{\ell,j} \\[.05in] f^{-1,1}_{\ell,j}  \end{array}\right)\quad &
\bF_{\ell j}^{0,1}=\left(  \begin{array}{c} 0 \\[.05in] f^{0,1}_{\ell,j} \end{array}\right)\quad &
\bF_{\ell j}^{1,1}=\left(  \begin{array}{c} f^{1,1}_{\ell,j} \\[.05in] -f^{1,1}_{\ell,j} \end{array}\right)\quad \\ \\
\bF_{\ell j}^{-1,0}=\left(  \begin{array}{c} f_{\ell,j}^{-1,0} \\[.05in] 0 \end{array}\right)\quad && 
\bF_{\ell j}^{1,0}=\left(  \begin{array}{c} -f_{\ell,j}^{1,0}  \\[.05in] 0 \end{array}\right)\quad \\ \\
\bF_{\ell j}^{-1,-1}=\left(  \begin{array}{c} f^{-1,-1}_{\ell,j} \\[.05in] f^{-1,-1}_{\ell,j}  \end{array}\right)\quad &
\bF_{\ell j}^{0,-1}=\left(  \begin{array}{c} 0 \\[.05in] f^{0,-1}_{\ell,j} \end{array}\right)\quad & 
\bF_{\ell j}^{1,-1}=\left(  \begin{array}{c} f^{1,-1}_{\ell,j} \\[.05in] -f^{1,-1}_{\ell,j} \end{array}\right), 
\end{array}
\end{equation}
with $f_{\ell j}^{mn}$ defined as
\begin{equation}
f_{\ell j}^{mn} = 
\left\{ 
\begin{array}{cl} 
\sigma_{i,j;n,m} \   \epsilon_{i,j;m,n}\  k_D a_{ss} & \quad\xtext{if} \quad (m,n) \in (\pm 1,\pm 1) \\
\sigma_{i,j;n,m} \phantom{.} \epsilon_{i,j;m,n} \phantom{.}  k_L a_{ss} & \quad\xtext{otherwise}.  \\
\end{array}
\right.
\end{equation}
The connectivity matrix, $\sigma_{i,j;m,n}$, is defined as
\begin{equation}
\label{connectivity_matrix}
\sigma_{\ell,j;n,m}=\left\{\begin{array}{ll}
1 & \quad \xtext{if sites}\  (\ell,j)\  \xtext{and} \  (\ell+m,j+n) \ \xtext{both contain atoms}\\
0 & \quad \xtext{otherwise}.
\end{array}
\right.
\end{equation}
and the misfit matrix, $\epsilon_{\ell,j;n,m}$, is defined as
\begin{equation}
\epsilon_{\ell,j;n,m}=\left\{\begin{array}{ll}
\epsilon_{gg} &  \quad \xtext{if sites}\ (\ell,j)\  \xtext{and} \  (\ell+m,j+n) \ \xtext{both contain Ge atoms}\\
\epsilon_{sg} &  \quad \xtext{if sites}\  (\ell,j)\  \xtext{and} \  (\ell+m,j+n) \ \xtext{contain a Ge and a Si atom}\\
0 & \quad \xtext{otherwise}
\end{array}
\right.
\end{equation}

\paragraph{Note :}The misfit for a Ge/Si system is $\epsilon_{gg}=0.04$ is known.
But our model demands a parameter $\epsilon_{sg}$ which is not clear. 
We have conducted simulations with both $\epsilon_{sg}=0.04$ and $\epsilon_{sg}=.02$. 
The first choice considers $a_{sg}=a_{gg}$ and the second considers $a_{sg}=0.5 a_{gg}$. 
The second choice seems appropriate if one considers the two species as hard spheres of 
different sizes.
Preliminary simulations with the first choice revealed quantitative differences
in the island sizes but no apparent qualitative differences in the nature of the instability. 
With the true nature of the interactions being unknown, we present only results
corresponding to $\epsilon_{sg}=0.02$.

\subsection{Interactions}
The atoms in the reference state experience forces from their neighbors as
given by Eqn. \ref{netforce}. 
The atoms become displaced from the reference position to
achieve mechanical equilibrium (force balance) and we denote
the  displacement of the atom at site $(\ell,j)$ as $ (u_{\ell j},
v_{\ell j})^T $.

Our aim is to calculate  $ (u_{\ell j} , v_{\ell j})^T $. 
We will assume that there is a $J$ such that the crystal is made of pure Si for $j \leq J$ 
and that all sites in this region are occupied. This means that
in the region $j \leq J$ the  misfit forces given by Eqn. \ref{netforce}
will be zero. Whereas these forces will be nonzero
for $j \geq J$;
this includes the deposited atoms and the regions where the deposited 
atoms have intermixed with the substrate atoms. 
We treat these two regions separately. 
In order to set a convenient language for the rest of this paper we will
refer to these regions as the film ($j\geq J$) and substrate ($j<J$), 
not to be confused with the original substrate and the deposited material.
Further we choose our indices such that $J = 0$ for convenience. 

\subsubsection{ Interactions in the Film Region}
For the film region corresponding to the indices $j\geq 0$, the force balance gives us

\begin{eqnarray}
0=F_{\ell j}
  &+& k_L\left(
           \sigma_{\ell,j;1,0} (u_{\ell+1,j}-u_{\ell j})
           +\sigma_{\ell,j;-1,0}(u_{\ell-1,j}-u_{\ell j})\right)\nonumber\\
   &+&\frac{k_D}{2}\left(
           \sigma_{\ell,j;1,1} (u_{\ell+1,j+1}-u_{\ell j})
           +\sigma_{\ell,j;-1,1}(u_{\ell-1,j+1}-u_{\ell j})\right)\nonumber\\
   &+&\frac{k_D}{2}\left(
           \sigma_{\ell,j;1,-1} (u_{\ell+1,j-1}-u_{\ell j})
           +\sigma_{\ell,j;-1,-1}(u_{\ell-1,j-1}-u_{\ell j})\right)\nonumber\\
   &+&\frac{k_D}{2}\left(
           \sigma_{\ell,j;1,1}  (v_{\ell+1,j+1}-v_{\ell j})
          -\sigma_{\ell,j;-1,1} (v_{\ell-1,j+1}-v_{\ell j})\right) \nonumber\\
   &+&\frac{k_D}{2}\left(
        -\sigma_{\ell,j;1,-1}  (v_{\ell+1,j-1}-v_{\ell j})
       + \sigma_{\ell,j;-1,-1}  (v_{\ell-1,j-1}-v_{\ell j})\right) \label{FX}
\end{eqnarray}
and
\begin{eqnarray}
0=G_{\ell j} 
  &+& k_L\left(
           \sigma_{\ell,j;0,1}(v_{\ell j+1}-v_{\ell j})+
           \sigma_{\ell,j;0,-1} (v_{\ell,j-1}-v_{\ell j})\right)\nonumber\\
   &+&\frac{k_D}{2}\left(
           \sigma_{\ell,j;1,1}(v_{\ell+1,j+1}-v_{\ell j})+
           \sigma_{\ell,j;-1,1}(v_{\ell-1,j+1}-v_{\ell j})\right)\nonumber\\
   &+&\frac{k_D}{2}\left(
          \sigma_{\ell,j;1,-1}(v_{\ell+1,j-1}-v_{\ell j})+
         \sigma_{\ell,j;-1,-1}(v_{\ell-1,j-1}-v_{\ell j})\right)\nonumber\\
   &+&\frac{k_D}{2}\left(
         \sigma_{\ell,j;1,1}(u_{\ell+1,j+1}-u_{\ell j})
       -\sigma_{\ell,j;-1,1}(u_{\ell-1,j+1}-u_{\ell j})\right) \nonumber\\
   &+&\frac{k_D}{2}\left(
        -\sigma_{\ell,j;1,-1}(u_{\ell+1,j-1}-u_{\ell j})+
         \sigma_{\ell,j;-1,-1}(u_{\ell-1,j-1}-u_{\ell j})\right) \label{FY}
\end{eqnarray}
where $\sigma_{\ell,j;n,m}$ is the connectivity matrix defined in Eqn.\ref{connectivity_matrix}.
This gives a linear system of equations that can be solved for $ (u_{\ell j}, v_{\ell j}) $. 
This system interacts with the substrate region. 
Note that the above relations at $j=0$ use $u_{\ell,-1},v_{\ell,-1}$
which are displacements in the substrate. 

\subsubsection{Interactions in the Substrate Region}
Now for the region $j < 0$ we first note that, 
this is a semi-infinite substrate. 
There are no forces arising from misfit in the substrate and requiring mechanical equilibrium gives us

\begin{eqnarray}
 0  = &\phantom{+}& k_L (u_{\ell+1,j} - 2 u_{\ell,j } + u_{\ell-1,j }) \nonumber\\
               &+& \frac{k_D}{2} (u_{\ell+1, j+1} + u_{\ell-1, j+1 }
                     + u_{\ell+1,j-1 } + u_{\ell-1,j-1 }-4u_{\ell j }) \nonumber\\
               &+&  \frac{k_D}{2} (v_{\ell+1,j+1}+v_{\ell-1,j-1}
                     - v_{\ell+1 j-1}-v_{\ell-1 j+1})
                     \label{eq:fbulk2D}
\end{eqnarray}
\begin{eqnarray}
  0  = &\phantom{+}& k_L (v_{\ell j+1 } - 2 v_{\ell j } + v_{\ell j-1 }) \nonumber \\
               & +& \frac{k_D}{2}(v_{\ell+1,j+1 } + v_{\ell-1,j+1 }
            + v_{\ell +1, j-1} + v_{\ell-1 ,j-1}- 4v_{\ell, j}) \nonumber \\
              & +&  \frac{k_D}{2} (u_{\ell+1 j+1 } + u_{\ell-1 j-1}
                  - u_{\ell+1 j-1 } - u_{\ell-1 j+1}).
                 \label{eq:gbulk2D}
\end{eqnarray}
This is a homogeneous system of equations that takes a non-trivial solution due to 
the displacement field of the film (at $j = 0$ ). 
This gives us the displacement field for $j < 0$ of a
relaxed substrate with given displacements at $j=0$.
The system in  Eqn. \ref{eq:fbulk2D} and \ref{eq:gbulk2D} needs 
to be solved for $j < 0$ given $(u_{\ell,0},v_{\ell,0})$. 
We do this by using a Fourier series in the $x$-direction:
\begin{equation}
\left(\begin{array}{cc} u_{\ell j}\\ v_{\ell j} \end{array}\right) =
\sum_{\xi=1}^M
\left(\begin{array}{cc}\hu_{\xi j}\\ \hv_{\xi j} \end{array}\right) e^{i\ell\xi}.\nonumber
\end{equation}
Now inserting this into Eqn. \ref{eq:fbulk2D} and \ref{eq:gbulk2D} we get

\begin{eqnarray}
  0  & = & 2 k_L \hu_j(\cos\xi-1)
               + k_D \left[(\hu_{\xi,j-1}+\hu_{\xi,j+1})\cos\xi  - 2\hu_{\xi j})\right.\nonumber \\
         &\phantom{=}&
               + i\left.(\hv_{\xi,j+1}-\hv_{\xi,j-1})\sin\xi \right]  \nonumber \\[-0.2cm]
 & & \label{eq:Fbulk} \\[-0.2cm]
  0  & = & 2 k_L (\hv_{\xi,j+1}-2\hv_{\xi j}+\hv_{\xi,j-1})
               + k_D\left[(\hv_{\xi,j-1}+\hv_{\xi,j-1})\cos\xi  -2\hv_{\xi,j}\right.\nonumber \\
         &\phantom{=}&
               + i\left.(\hu_{\xi,j+1}-\hu_{\xi,j-1})\sin\xi\right]. \nonumber
\end{eqnarray}

We look for a solution for the above system in the form
\[
   \hat{u}_{\xi j} = \alpha^j\hat{u}_{\xi,0}, \quad \hat{v}_{\xi,j} = \alpha^j\hat{v}_{\xi,0}.
\]
and obtain a linear homogeneous system of the form
\begin{equation}
\label{eq:uhat}
\mathbf{\Omega} ( \alpha ) \left( \begin{array}{c}  \hat{u}_{\xi,0} \\ \hat{v}_{\xi,0} \end{array} \right) = 0.
\end{equation}
The above system admits non trivial solution if
\[
P(\alpha) := det[\mathbf{\Omega}(\alpha)]=0
\]
giving raise to a polynomial equation in $\alpha$. 
The polynomial $P(\alpha)$ is of degree 4 and has roots of the form 
$(\alpha_1,\alpha_2,1/\alpha_1,1/\alpha_2)$ 
where $|\alpha_1 | , |\alpha_2 | > 1$ \cite{russo_smereka_3D}. 
Since we are interested in solutions $\hat{u}_j = \alpha^j\hat{u}_0, \quad \hat{v}_j = \alpha^j\hat{v}_0$ 
that decay as $ j \to -\infty$ we pick the two roots $\alpha_1, \alpha_2$.
\begin{equation}
\left(\begin{array}{cc} \hu_{\xi,j} \\ \hv_{\xi,j}\end{array}\right)= Q(j)Q^{-1}(0)
\left(\begin{array}{cc} \hu_{\xi,0} \\ \hv_{\xi,0}\end{array}\right)\label{DUFFT}
\end{equation}
where $Q(j)$ is the invertible $2 \times 2$ matrix
given by
\[
\left( \vec{r_1}\alpha_1^j \quad \vec{r_2}\alpha_2^j \right).
\]
Finally, $\vec{r_p}$ and $\alpha_p$ are the eigenvectors and eigenvalues
that arise when solving the discrete equation.
The solution of Eqn. \rf{eq:fbulk2D} and \rf{eq:gbulk2D} is then
given by
\begin{equation}
\left(\begin{array}{cc} u_{\ell,j} \\ v_{\ell,j}\end{array}\right)=
    \sum_{\xi=1}^M Q(j)Q^{-1}(0)
\left(\begin{array}{cc}\hu_{\xi 0}\\ \hv_{\xi 0} \end{array}\right) e^{i\ell\xi}
\quad\xtext{where}\quad
\left(\begin{array}{cc}\hu_{\xi 0}\\ \hv_{\xi 0} \end{array}\right) =
  \frac1M \sum_{\xi=1}^M 
\left(\begin{array}{cc} u_{\ell,0} \\ v_{\ell,0}\end{array}\right)e^{-i\xi\ell} \label{FFT}
\end{equation}
A three dimensional version  of this was presented in \cite{russo_smereka_3D} and
similar approach was presented in \cite{Caflisch2}.
 
\section{Elastic Energy of the Crystal.}
The elastic energy can be computed using the displacement field calculated 
by the algorithm in the previous section; it is
\begin{equation}
\label{eqn:elastic_energy_sum}
W = \sum_{\xtext{all springs}} \frac{1}{2} k_{spring} \delta^2
\end{equation}
where $k_{spring} = k_L, k_D$ depending on whether the spring is a diagonal or lateral spring,
and $\delta$ is the change in spring length with respect to natural spring length. We
rewrite $W$ as
\begin{equation}
W = W_f + W_s \label{Wtot}
\end{equation}
where
\[
W_f = \sum_{j > 0} \frac{1}{2} k_{spring} \delta^2
\quad\xtext{and}\quad
W_s = \sum_{j \le 0} \frac{1}{2} k_{spring} \delta^2
\]
where $\sum_{j > 0}$ indicates summing all springs connected to atoms located
at sites with $j > 0$, $\sum_{j \le 0}$ is defined in a similar way. We can
write the expression of $W_f$ as a sum over atoms:
\begin{equation}
W_f = \frac12\sum_{j \ge 0} w_{i,j}\label{WF}
\end{equation}
where $w_{i,j}$ is the energy of all the springs connected to the
atom at site $(i,j)$. The expression for $w_{i,j}$ is given at the end of this section.
Since there are no forces arising from misfit in the
substrate we can write the expression for the elastic energy of the substrate
\[
W_{s} = -\frac{1}{2} {\bf u}_s^{T} {\bf A}_{s} {\bf u}_{s}
\label{eq:subs_energy}
\]
where ${\bf A}_{s}$ is the non-positive infinite dimensional matrix
representing the interaction of the atoms in the substrate and 
${\bf u}_{s}$ is the vector representing their displacements. 
It is convenient to decompose
${\bf u}_{s}$ into the displacements of the atoms at $j=0$ and those
for $j < 0$:
\[
{\bf u}_s=\left(\begin{array}{ll} {\bf u}_0 \\
{\bf u}_{j<0} 
\end{array}\right)
\]
Since atoms in the substrate below the first layer produce no
net force on each other it follows 
\[
{\bf A}_{s} 
\left(\begin{array}{ll} {\bf u}_0 \\
{\bf u}_{j<0} \end{array}\right)=
\left(\begin{array}{ll} {\bf f}_0\\ 
0 \end{array}\right) 
\]
where ${\bf f}_0$ is the force on the top layer of the
substrate atoms due to the substrate atoms. If we use the above expression
in Eqn. \ref{eq:subs_energy} we find
\[
W_s =  -\frac{1}{2} {\bf f}_0\cdot {\bf u}_0 
\]
which can be written as
\begin{equation}
W_s = -\frac12\sum_{\ell=1}^M (u_{\ell,0}f_{\ell}+v_{\ell,0}g_{\ell})\label{WS}
\end{equation}
where
\[
f_{\ell} = k_L (u_{\ell-1, 0} - 2 u_{\ell, 0} + u_{\ell+1}) 
             +  \frac{k_D}{2} (u_{\ell-1,-1} - 2u_{\ell, 0}+ u_{\ell+1,-1} )
                +   \frac{k_D}{2} (v_{\ell-1, -1} -v_{\ell+1, -1})
\]
and
\[
g_{\ell} = k_L (-v_{\ell, 0 } +  v_{\ell, -1 } ) 
                + \frac{k_D}{2}(v_{\ell-1,-1 } - 2v_{\ell, 0} + v_{\ell+1,-1}) 
               +  \frac{k_D}{2} (u_{\ell-1, -1 } - u_{\ell+1,-1}).
\]
Therefore the total energy of the crystal is given by 
Eqn. (\ref{Wtot}) along with Eqns. (\ref{WF}) and (\ref{WS}).
It should be pointed out that this expression could have also been obtained
using summation by parts.
Finally, as
promised we present the expression for $w_{i,j}$
\begin{equation}
w_{i,j}=w^{xx}_{i,j}+w^{yy}_{i,j}+2w^{xy}_{i,j} \label{dis_enden}
\end{equation}
with
\begin{eqnarray*}
w^{xx}_{i j}
 &=& \frac{k_L}{2}\left(
      \sigma_{i,j;1,0}(u_{i+1,j}-u_{i, j}-\ssx)^2+
      \sigma_{i,j;-1,0}(u_{i -1,j}-u_{i, j}+\ssx)^2\right)\\
 &+&\frac{k_D}{4}\left(
     \sigma_{i,j;1,1} (u_{i+1,j+1}-u_{i, j}-\ssx)^2+
     \sigma_{i,j;-1,-1} (u_{i-1,j-1}-u_{i, j}+\ssx)^2\right.\\
    &&\left.\phantom{pppp}+\sigma_{i,j;1,-1} (u_{i+1,j-1}-u_{i, j}-\ssx)^2+
     \sigma_{i,j;-1,1} (u_{i-1,j+1}-u_{i, j}+\ssx)^2\right),\\
\end{eqnarray*}
\begin{eqnarray*}
w^{yy}_{i j}
 &=& \frac{k_L}{2}\left(
      \sigma_{i,j;0,1}(v_{i,j+1}-v_{i, j}-\ssy)^2+
      \sigma_{i,j;0,-1}(v_{i,j-1}-v_{i,j j}+\ssy]^2\right)\\
 &+&\frac{k_D}{4}\left(
     \sigma_{i,j;1,1} (v_{i+1,j+1}-v_{i, j}-\ssy)^2+
     \sigma_{i,j;-1,-1} (v_{i-1,j-1}-v_{i, j}+\ssy)^2\right.\\
    &&\left.\phantom{pppp}+\sigma_{i,j;1,-1} (v_{i+1,j-1}-v_{i,j}+\ssy)^2+
     \sigma_{i,j;-1,1} (v_{i-1,j+1}-v_{i, j}-\ssy)^2
\right),\\
\end{eqnarray*}
\begin{eqnarray*}
w^{xy}_{i j}& =& \frac{k_D}{4}\left(
     \sigma_{i,j;-1,-1} (u_{i-1,j-1}-u_{i, j}+\ssx)
                           (v_{i-1,j-1}-v_{i, j}+\ssy)
     \right.\\
     &&\left.\phantom{pppp}+
     \sigma_{i,j;1,1}  (u_{i+1,j+1}-u_{i, j}-\ssx)
                        (v_{i+1,j+1}-u_{i, j}-\ssy)\right.\\
     &&\phantom{pppp}\left.-\sigma_{i,j;1,-1}
   (u_{i+1,j-1}-u_{i, j}-\ssx)(v_{i+1,j-1}-v_{i, j}+\ssy)\right.\\
     &&\phantom{pppp}\left.-\sigma_{i,j;-1,1}
(u_{i-1,j+1}-u_{i, j}+\ssx)(v_{i-1,j+1}-v_{i, j}-\ssy)
\right),\\
\end{eqnarray*}
where
\begin{equation}
   \ssx = \left\{\begin{array}{cl}
                 a_{gs}-a_{ss} & {\quad\xtext{Ge-Si bonds}} \\
                 a_{gg}-a_{ss} & {\quad\xtext{Ge-Ge bonds}} \\
                0 & {\quad\xtext{for Si-Si bonds}}
             \end{array}
             \right.\label{SSX} \,
\end{equation}

\section{Multigrid Fourier Algorithm}
In this section we describe a multigrid Fourier algorithm to solve the linear system, 
derived in the previous section, for the displacement field. This algorithm
is quite similar to the one presented in \cite{russo_smereka_2D} but there
are differences, first it was extended to a multicomponent system, second
the incorporation of the substrate was simplified as was the coarse graining
procedure. In \cite{russo_smereka_2D} the substrate was removed and replaced
by effective forces.  Unfortunately standard SOR (the method of Successive
over Relaxation) is unstable when this is done
and the resulting system had to be under relaxed to stabilize it. 
Here the substrate is handled in more seem less way
by using Eqn. \ref{FFT} to supply the boundary conditions. As we shall see this
formulation coarse grains nicely resulting in an efficient algorithm.

Multigrid is an efficient method for solving linear
systems $\mathbf{Au}+\bF=0$ where $\mathbf{A}$ comes from the
discretization of a partial differential equation. 
A good introduction to the idea of a multigrid algorithm 
can be found in the book by Briggs \cite{briggs}.
Two important tools needed to implement multigrid algorithm 
are coarse graining, $\sf CF$, and prolongation operators, $\sf FC$.
The operator $\sf CF$ takes data from fine grid to a coarse grid;
if $\bf u$ is a vector of length $N$ then $\sf{CF}\bf u$
will be vector of length $N/2$. The operator $\sf FC$ takes data
from a coarse grid  to fine grid, if $\bf u$ is a vector of length $N$ then $\sf{FC}\bf u$
is vector of length $2N$. If $\bf u$ is smooth then
$\bf u \approx {\sf FC}({\sf CF}\bf u )$.
Naturally there are many different ways to formulate these operators
but typically coarse-graining operators are based on averaging and
the prolongation operators are based on interpolation.
In addition, a multigrid algorithm needs a course grained version
of $\bf A$, denoted as ${\bf A}^{(2)}$. If $\bf A$ is $N\times N$
then ${\bf A}^{(2)}$ will be $N/2\times N/2$. One natural way
to construct ${\bf A}^{(2)}$ is to consider the discretization
on a coarser grid.

Our goal is to solve $\mathbf{Au}+\bF=0$. Suppose we have an 
an approximate solution, $\mathbf{u}_a$. To assess its accuracy we
can compute the residual: $\mathbf{r}=\mathbf{Au}_a+\bF$. The
relationship between the residual and the error, ${\bf e}={\bf u}- {\bf u}_a$,
is 
\begin{equation}
\mathbf{Ae}+\bf r=0 \label{AER}
\end{equation}
One could solve the above equation for $\bf e$ and
then determine $\bf u$ using ${\bf u}={\bf u}_a+{\bf e}$,
at first sight that would appear as much work as solving the
original system.  The crucial observation used  by multigrid methods
is that if an iterative solver such as Jacobi 
Gauss Seidal or  (SOR) is used to obtain the approximate solution  then
the residual, $\mathbf{r}$, and the error, $\mathbf{e}$, are known to be smooth. Indeed
this is why these methods converge so slowly. 
Since $\mathbf{r}$ is
smooth  little information is lost because of coarse-graining. For this reason
multigrid methods replace Eqn. \ref{AER} by
\begin{equation}
{\bf A}^{(2)}{\bf e}^{(2)}+{\bf r}^{(2)}=0 \label{AER2}
\end{equation}
Once this smaller system is solved the solution then updated using
${\bf u}_a^{new}={\bf u}_a+{\sf FC}{\bf e}^{(2)}$. One then
applies SOR or Jabobi on the fine scale using ${\bf u}_a^{new}$
as a guess. This procedure is repeated until the residual is
sufficiently small. This represents a two level multigrid
scheme. In most implementations this procedure is extended to
many levels.

Implementation of a multigrid method for the system given by
Eqns. \ref{FX}-\ref{FY} could be accomplished with a standard multigrid method if
the film profile was flat and there was no artificial boundary
condition.  The main difficulty is the formulation of coarse-grained versions
of Eqns. \ref{FX}, \ref{FY}, and \ref{FFT}. One approach is to
use algebraic multigrid methods as in \cite{CafMG}. The approach
outlined here was presented in \cite{russo_smereka_2D} in which
the problem was defined on a rectangular domain using fictitious atoms. 

\subsection{Fictitious atoms}
\begin{figure}[!ht]
\begin{center}
\includegraphics[width=4in]{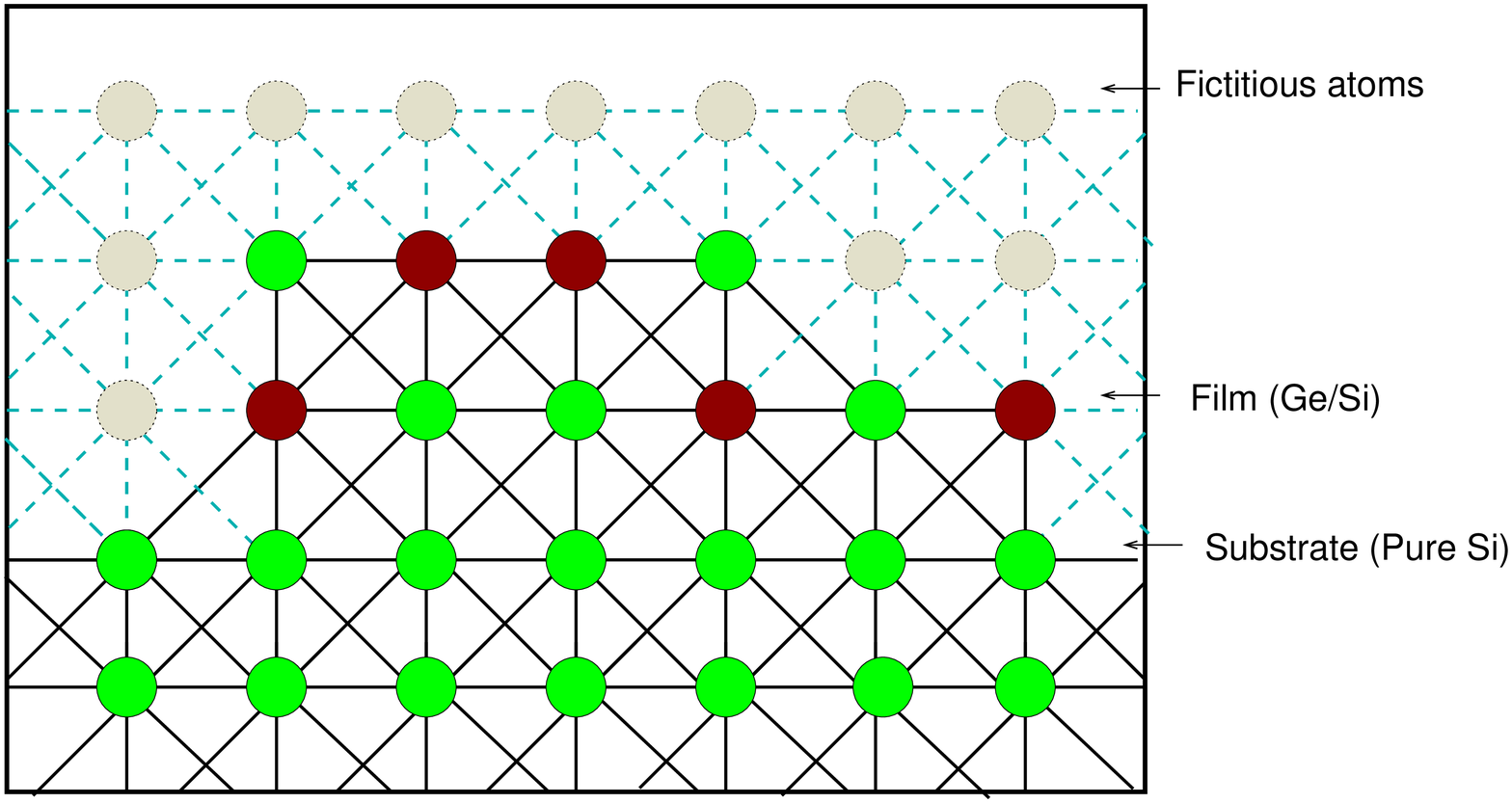}
\end{center}
\caption{\sf \small Computational domain, with real and fictitious atoms.
The red circles represent the Ge atoms and the green circles the Si atoms.
The null interaction with fictitious atoms is represented by dashed lines.
}
\label{fig:fictitious}
\end{figure}

We start by defining a rectangular domain. Let $j_{max}$ be the vertical coordinate of 
the highest atom. 
Then we define our domain of computation to be 
\[
\Omega=\{(i,j): \quad 1\le i\le M, \quad 0\le j \le N \}.
\]
where $M$ is the period of the lattice in the horizontal direction and $N>j_{max}$ is an integer. 
Some of these sites are occupied and others are not. 
All sites in $\Omega$ that are not occupied by real atoms are called fictitious atoms. 
This is illustrated in the figure (Fig: \ref{fig:fictitious} ).
The system of equations in \rf{FX} and \rf{FY} are now extended to fictitious atoms, 
simply by setting the connectivity matrix to zero at these sites. 
This is done by defining the \emph{site atom density}

\begin{equation}
   p_{i,j} = \left\{\begin{array}{cl}
                1 & \quad {\rm if} \> (i,j) \> \mbox{is a real atom} \\
                0 & \quad {\rm if} \> (i,j) \> \mbox{is a fictitious atom}
             \end{array}
             \right. \, .\label{ATOMDEN}
\end{equation}
With these definitions, the connectivity matrix can be written as
\begin{equation}
   \sigma_{i,j;m,n} = p_{i,j} p_{i+m,j+n}.
   \label{eq:connectivity}
\end{equation}
In order to simplify the notation we introduce the following spring strength matrices
\[
\begin{array}{llllll}
 K^x_{-1,-1}  & = \  k_D/2  &\quad K^x_{0,-1}  &= \  0 &\quad K^x_{1,-1} & =\   k_D/2\\[.05in]
K^x_{-1, 0} & = \  k_L    &\quad K^x_{0, 0}  &= \  0 & \quad K^x_{1, 0} &=\   k_L\\[.05in]
K^x_{-1, 1}  &= \  k_D/2 & \quad K^x_{0, 1}& = \  0 & \quad K^x_{1, 1} & =\   k_D/2
\end{array}
\]
and
\[
\begin{array}{llllll}
K^y_{-1,-1} & = \  k_D/2 &\quad K^y_{0,-1} & = \  k_L &  \quad K^y_{1,-1} & = \ k_D/2\\[.05in]
K^y_{-1, 0} & = \ 0     & \quad K^y_{0, 0} & = \  0   &  \quad K^y_{1, 0} & = \  0\\[.05in]
K^y_{-1, 1} & = \ k_D/2  & \quad K^y_{0, 1} & = \  k_L &  \quad K^y_{1, 1} & = \  k_D/2.
\end{array}
\]
Using these matrices we can rewrite (\ref{FX}) and (\ref{FY}) as:
\begin{equation}
          \sum_{m,n=-1}^1 K^x_{mn}\sigma_{\ell j; m n}
          \left(
               [u_{\ell+m,j+n}-u_{\ell j}]+m n
               [v_{\ell+m,j+n}-v_{\ell j}]
          \right) - F_{\ell j} =0
\label{FXC}
\end{equation}

\begin{equation}
          \sum_{m,n=-1}^1K^y_{mn}\sigma_{\ell j; m n}
          \left(
               [v_{\ell+m,j+n}-v_{\ell j}]+m n
               [u_{\ell+m,j+n}-u_{\ell j}]
          \right) - G_{\ell j} =0.
          \label{FYC}
\end{equation}

\subsection{Coarsening and Prolongation Operations}

A key ingredient in the multigrid technique are the coarsening and prolongation 
operations used to map data between coarser and finer grids.  We let $L = 1,2,\ldots L_g$ 
denote the scale of the grid.  $L=1$ denoted the finest scale and $L_g$
the coarsest. In our computations we will choose the number of grid
points in the horizontal direction on the finest scale to be $M=2^P$ 
where $P$ is an integer. Clearly then, $L_g \le P$.  
The number grid points in the horizontal
direction for the other levels are
\begin{equation}
M_{L+1} = \frac{M_L}{2} \qquad L=1,2,\ldots, L_g-1 
\label{MM}
\end{equation}
where $M_1=M$. In the vertical direction, we let $N$ denoted the number of grid
points on the finest scale. The number of grid points 
for the other levels is given by
\begin{equation}
N_{L+1} =\max( \integer{\frac{N_{L}+1}{2}}, 1) \qquad L=1,2,\ldots,  L_g-1
\label{NN}
\end{equation}
where $\integer{\cdot}$ represents the integer part. 

\begin{figure}[!ht]
\begin{center}
\includegraphics[width=5in]{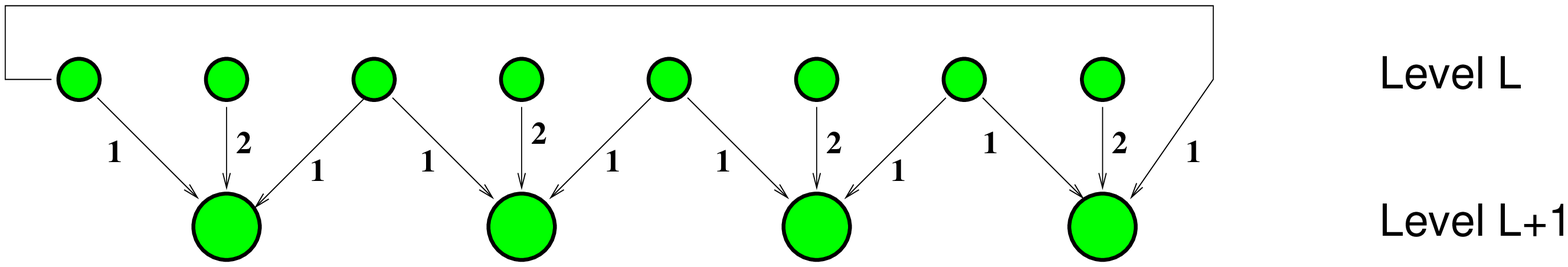}
\end{center}
\caption{\sf \small Coarsening operation in the horizontal direction 
with even number of grid points.
The same relative weights are used in the prolongation, when interpolating
values from the coarse to the fine grid. 
}
\label{fig:coarse_x_even}
\end{figure}

\subsubsection{Operators in horizontal direction}
In the case of the horizontal direction we have periodic boundary
conditions and the
number of grid points will be even
for all levels.  The coarse graining procedure we use
is displayed schematically on Fig. \ref{fig:coarse_x_even} 
and can be expressed as
\begin{equation}
{\bf q}^{L+1}= C_x{\bf q}^L
\label{FCX}
\end{equation}
where ${\bf q}^L$ is a quantity defined on the fine grid,
${\bf q}^{L+1}$ is defined on the coarse grid,
and $C_x$ is the $M_{L+1}\times M_L$ matrix given by
\[
C_x= \frac14\left(\begin{array}{ccccccccccccccccccc}
1 & 2 & 1&  &  &     &     &     &  &  &      &     &     &  &  &  &    &    \\[.0in]
  &  &  1& 2& 1&     &     &     &  &  &      &     &     &  &  &  &    &    \\[.0in]
  &  &   &  &1 &   2 &   1 &     &  &  &      &     &     &  &  &  &    &    \\[.0in]
  &  &  &  &  &\cdot&\cdot&\cdot&  &  &      &     &     &  &  &  &    &    \\[.0in]
  &  &  &  &  &     &     &     & 1&2 &    1 &    &     &  &  &  &    &    \\[.0in]
  &  &  &  &  &     &     &     &  &  & \cdot&\cdot&\cdot&  &  &  &    &    \\[.0in]
  &  &  &  &  &     &     &     &  &  &      &     &     &  &  &  &   &     \\[.0in]
 1& 0& \cdot&  \cdot&  \cdot&  &     &     &     &  &  &      &     &     &  &  &  1& 2&\\[.0in]
\end{array}\right)
\]
The operation to take variables from the coarse grid to
the fine grid is simply
\begin{equation}
{\bf q}^{L}= P_x{\bf q}^{L+1} \quad\xtext{where}\quad P_x=2C_x^T
\label{CFX}
\end{equation}

\subsubsection{Operators in vertical direction}

\begin{figure}[!ht]
\begin{center}
\includegraphics[width=5in]{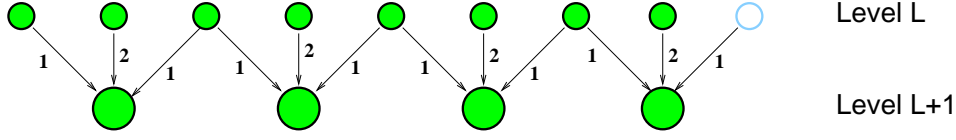}
\end{center}
\caption{\sf \small Coarsening operation in the vertical direction for an even number of grid points
on the fine scale.
The blue circles represent the grid outside the computational domain where the data is assumed to be zero.}
\label{fig:coarse_y_even}
\end{figure}

\begin{figure}[!ht]
\begin{center}
\includegraphics[width=5in]{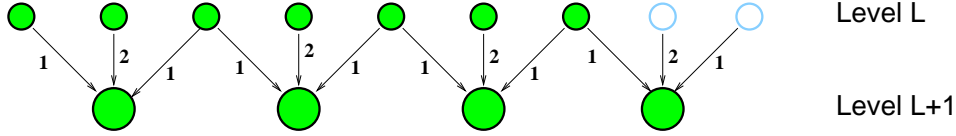}
\end{center}
\caption{\sf \small Coarsening operation in the vertical direction for an odd number of grid points
on the fine scale}
\label{fig:coarse_y_odd}
\end{figure}

The situation in the vertical direction is slightly different mainly because
the number of grid points, $N_L$, can be either even or odd. Our coarse
graining strategy is shown in Fig. \ref{fig:coarse_y_even}  and Fig. \ref{fig:coarse_y_odd}.
We can write the coarse graining and prolongation operations as follows
\[
{\bf q}^{L+1}= C_y{\bf q}^{L} \quad\xtext{and}\quad 
{\bf q}^{L}= P_y{\bf q}^{L+1}
\]
The matrices $C_y$ and $P_y$ are defined differently for even and odd
$N_L$.
For $N_L$ even we have
\[
\begin{array}{c}
C_y= \frac14\left(\begin{array}{ccccccccccccccccccc}
1 & 2 & 1&  &  &     &     &     &  &  &      &     &     &  &  &  &    &    \\[.0in]
  &  &  1& 2& 1&     &     &     &  &  &      &     &     &  &  &  &    &    \\[.0in]
  &  &   &  &1 &   2 &   1 &     &  &  &      &     &     &  &  &  &    &    \\[.0in]
  &  &  &  &  &\cdot&\cdot&\cdot&  &  &      &     &     &  &  &  &    &    \\[.0in]
  &  &  &  &  &     &     &     & 1&2 &    1 &    &     &  &  &  &    &    \\[.0in]
  &  &  &  &  &     &     &     &  &  & \cdot&\cdot&\cdot&  &  &  &    &    \\[.0in]
  &  &  &  &  &     &     &     &  &  &      &     &     &  & 1 & 2 & 1&     \\[.0in]
  &  &  &  &  &  &     &     &     &  &  &      &     &     &  &  &  1& 2&\\[.0in]
\end{array}\right)
\quad\xtext{and}\quad \\
P_y= \frac12\left(\begin{array}{cccccccc}
2 &   &   &   &   &   &   &   \\
2 &   &   &   &   &   &   &   \\
1 & 1 &   &   &   &   &   &   \\
  & 2 &   &   &   &   &   &   \\
  & 1 & 1 &   &   &   &   &   \\
  &   & 2 & \cdot  &   &   &   &   \\
  &   & 1 & \cdot  &   &   &   &   \\
  &   &   & \cdot &   &   &   &   \\
  &   &   &   &   &  1 &   &   \\
  &   &   &   &   &  2 &  &   \\
  &   &   &   &   &  1 & 1 &   \\
  &   &   &   &   &     & 2 &   \
\end{array}\right),
\end{array}
\]
whereas for $N_L$ odd we have
\[
\begin{array}{c}
C_y= \frac14\left(\begin{array}{ccccccccccccccccccc}
1 & 2 & 1&  &  &     &     &     &  &  &      &     &     &  &  &  &    &    \\[.0in]
  &  &  1& 2& 1&     &     &     &  &  &      &     &     &  &  &  &    &    \\[.0in]
  &  &   &  &1 &   2 &   1 &     &  &  &      &     &     &  &  &  &    &    \\[.0in]
  &  &  &  &  &\cdot&\cdot&\cdot&  &  &      &     &     &  &  &  &    &    \\[.0in]
  &  &  &  &  &     &     &     & 1&2 &    1 &    &     &  &  &  &    &    \\[.0in]
  &  &  &  &  &     &     &     &  &  & \cdot&\cdot&\cdot&  &  &  &    &    \\[.0in]
  &  &  &  &  &     &     &     &  &  &      &     &     &  &  & 1& 2&  1   \\[.0in]
  &  &  &  &  &  &     &     &     &  &  &      &     &     &  &  &  & 1&\\[.0in]
\end{array}\right)
\quad\xtext{and}\quad \\
P_y= \frac12\left(\begin{array}{cccccccc}
2 &   &   &   &   &   &   &   \\
2 &   &   &   &   &   &   &   \\
1 & 1 &   &   &   &   &   &   \\
  & 2 &   &   &   &   &   &   \\

  & 1 & 1 &   &   &   &   &   \\
  &   & 2 & \cdot  &   &   &   &   \\
  &   & 1 & \cdot  &   &   &   &   \\
  &   &   & \cdot &   &   &   &   \\
  &   &   &   &   &  1 &   &   \\
  &   &   &   &   &  2 &   &   \\
  &   &   &   &   &  1 & 1 &   \\
\end{array}\right).
\end{array}
\]

\noindent
{\bf Remark.} It should be pointed out that $P_y$ is not quite equal to $2C_y^T$ as was the case
in the $x$ direction. In \cite{russo_smereka_2D} it was incorrectly asserted that
their prolongation operator, $P$,   could be written as twice the transpose of
the coarsening operator.
 
\subsubsection{Coarsening and Prolongation Operators in Two Dimensions}
We construct our coarse from fine operator, 
${\sf{CF}}$, and the fine from coarse operator ${\sf{CF}}$,
as the tensor product of the one dimensional
operators. Namely
\[
{\sf{CF}}=C_x\otimes C_y 
\quad\xtext{and}\quad {\sf{FC}}=P_x\otimes P_y
\]
Therefore if
${\bf q}^L$ is a quantity defined on the grid for
level $L$  then we map to the coarser grid of level
$L+1$ using
\[
{\bq}^{L+1} = {\sf{CF}}({\bq}^L)
\]
In an analogous way we can map ${\bf q}^L$ to
finer grid using
\[
{\bq}^{L-1} = {\sf{FC}}({\bq}^L)
\]

\subsubsection{Coarse-grained interactions}
Here it will be outlined how to coarse-grain Eqs. (\ref{FXC}) and (\ref{FYC}).
They will be written in terms of the coarse grained values of the 
\emph{site atom density}. The site atom density, $p_{i,j}$, given
by (\ref{ATOMDEN}), is coarsened for all
levels $L=2,\ldots L_g$ using $\sf CF$. The coarse-grained site atom density
is used to coarse-grain the connectivity matrix as follows
\[
\sigma^L_{i,j,m,n} = \sqrt{p^L_{i,j}}\sqrt{p^L_{i+m,j+n}},
\]
the geometric mean of the site atom densities at the two sites.
The coarse-grained version of Eqs. (\ref{FXC}) and (\ref{FYC})
are written as
\begin{eqnarray}
       2^{2-2L} \sum_{m,n=-1}^1
       c_{\ell j;mn}^L K^x_{mn}\left(
               u_{\ell+m,j+n}^L-u_{\ell j}^L+m n
               (v_{\ell+m,j+n}^L-v_{\ell j}^L)\right)  +F^L_{\ell j} =0\label{FXL}\\
      2^{2-2L} \sum_{m,n=-1}^1
          c_{\ell j;mn}^L K^y_{mn} \left(v_{\ell+m,j+n}^L-v_{\ell j}^L+m n
               (u_{\ell+m,j+n}^L-u_{\ell j}^L)\right)  + G^L_{\ell j}=0\label{FYL}
\end{eqnarray}
The factor $2^{2-2L}$ is typical of coarsened elliptic equations. 
This can be interpreted as the weakening of springs as several springs are
replaced by a single spring.
The boundary conditions at $j=-1$ are given by (\ref{FFT}) on the corresponding
grid level. In other words
\begin{equation}
\left(\begin{array}{cc} u^L_{\ell,j} \\ v^L_{\ell,j}\end{array}\right)=
    \sum_{\xi=1}^{M_L} Q(j)Q^{-1}(0)
\left(\begin{array}{cc}\hu_{\xi 0}\\ \hv_{\xi 0} \end{array}\right) e^{i\ell\xi}
\quad\xtext{where}\quad
\left(\begin{array}{cc}\hu_{\xi 0}\\ \hv_{\xi 0} \end{array}\right) =
  \frac{1}{M_L} \sum_{\xi=1}^{M_L}
\left(\begin{array}{cc} u_{\ell,0} \\ v_{\ell,0}\end{array}\right)e^{-i\xi\ell} \label{FFTL}
\end{equation}
The system of equations given by Eqs. (\ref{FXL}), (\ref{FYL}), and (\ref{FFTL}), 
will for sake of convenience, be written as
as
\begin{equation}
{\sf A}^L{\bu}^L+{\bF}^L=0.\label{SYSL}
\end{equation}
\noindent

\subsection{Successive Over Relaxation}\label{SSOR}

In our multigrid algorithm we shall use the method of successive over relaxations 
to generate approximate solutions.  To explain our implementation
it is useful to rewrite these equations as
\begin{equation}
   \left(\begin{array}{cc} a_{xx} & a_{xy} \\ a_{yx} & a_{xx} \end{array}\right)
   \left(\begin{array}{c} u_{\ell j}^L \\ v_{\ell j}^L\end{array}\right)
   =
   \left(\begin{array}{c} c_x \\ c_y \end{array}\right)
   -\left(\begin{array}{c} F_{\ell j}^L \\
   G_{\ell j}^L\end{array}\right),\label{eq:RF}
\end{equation}
where
\[
\begin{array}{ll}
\ds a_{xx}=2^{2-2L}  \sum_{m,n=-1}^1 c_{\ell j;mn} K^x_{mn}, \quad &
\ds a_{xy}=2^{2-2L} \sum_{m,n=-1}^1 c_{\ell j;mn} K^x_{mn} mn,  \\
\ds a_{yx}=2^{2-2L}  \sum_{m,n=-1}^1 c_{\ell j;mn} K^y_{mn}mn, \quad &
\ds a_{yy}=2^{2-2L} \sum_{m,n=-1}^1 c_{\ell j;mn} K^y_{mn},
\end{array}
\]
\[
c_x=2^{2-2L} \sum_{m,n=-1}^1
c_{\ell j;mn} K^x_{mn}\left( u_{\ell+m,j+n}^L+m n
               v_{\ell+m,j+n}^L\right),
\]
and
\[
c_y=2^{2-2L} \sum_{m,n=-1}^1
          c_{\ell j;mn} K^y_{mn} \left(v_{\ell+m,j+n}+m n
               u_{\ell+m,j+n}\right).
\]
We have omitted writing the explicit dependence of the $a$ and $c$ coefficients on $\ell$ and $j$. 
The relaxation scheme for the above system is based on treating the right hand side as known.
Let us denote by $((u_{\ell j}^L)^k, (v^L_{\ell j})^k)^T$ the
value of the displacement at iteration $k$. 
Then  the solution at the next iteration,
$((u_{\ell j}^L)^{k+1}, (v^L_{\ell j})^{k+1})^T$ the
is computed by relaxing system \rf{eq:RF} by one SOR iteration. 
The SOR relaxation is performed as
\begin{eqnarray}
   (u_{\ell,j}^L)^* & = & (c_x^k - F^L_{\ell j} - a_{xy} (v^L_{\ell j})^k)
   / a_{xx} \label{eq:unew}\\
  (u^L_{\ell j})^{k+1} & = & \omega (u^L_{\ell j})^* + (1-\omega) (u^L_{\ell j})^k\nonumber\\
   (v^L_{\ell,j})^* & = & (c_y^k - G^L_{\ell j} - a_{yx} (u^L_{\ell j})^{k+1})
   / a_{yy} \label{eq:vnew}\\
 (v^L_{\ell j})^{k+1} & = & \omega (v^L_{\ell j})^* + (1-\omega) (v^L_{\ell j})^k\nonumber
\end{eqnarray}
where the superscript $k$ on $c_x$ and $c_y$
indicates that these terms are evaluated using the displacements
at the $k$th iterate.

\subsection{ Multigrid V-cycle Implementation }
In our computations we implement the multigrid algorithm using a
standard V-cycle which, for sake of completeness, we will now
describe. The V-cycle starts with the following steps:

\medskip
\noindent{\it Precomputation}
\begin{enumerate}
\item Compute ${\bF}^1$ using (\ref{netforce}).
\item Compute $N_L$ and $M_L$ for $L=2$ to $L_g$ using (\ref{NN}) and (\ref{MM}).
\item Coarse-grain the site atom density:  $p^{L+1} = {\sf CF}( p^{L}) $  for $L=2$ to $L_g$
\item Compute the connectivity matrix: ( $\sigma^L_{i,j,m,n}$ ) for $L=2$ to $L_g$
\item Initialize first guess for ${\bu}^1_{guess}$ 
      (usually ${\bu^1_{guess}}=0$ or an existing field).
\end{enumerate}

\bigskip

\noindent{\it V-cycle}

\noindent
For $L=1$ to $L_g-1$ do the following

\hspace{.2in} Relax ${\sf A}^L{\bu}^L + {\bF}^L =0 $ for $\eta$ steps ($\eta =2$ in our calculations) 

\hspace{.2in} Compute residual ${\br^L}= {\bF}^L + {\sf A}^L{\bu}^L$

\hspace{.2in} Coarse Grain the residual: ${\br}^{L+1}={\sf CF}({\br}^L) $

\hspace{.2in} Set ${\bF}^{L+1}={\br}^{L+1}$

\hspace{.2in} Set ${\bu}^{L+1}= 0 $ (This is the initial condition for the next relaxation)

\noindent
End Loop

\vspace{.1in}

\noindent
Solve ${\sf A}^{L_g}{\bu}^{L_g} + {\bF}^{L_g}=0$ by relaxation

\vspace{.1in}

\noindent
For $L=L_g-1$ to $1$ do the following

\hspace{.2in} Prolong the solution on the coarse mesh: ${\be}^{L}={\sf FC}({\bu}^{L+1}) $

\hspace{.2in} Let ${\bu}^{L}_{guess}={\be}^L+{\bu}^L$

\hspace{.2in} Relax ${\sf A}^L{\bu}^L +{\bF}^L=0$ with initial guess ${\bu}^L_{guess}$ for $\eta$ steps

\noindent
End Loop

\medskip
The accuracy of the solution is measured using the $L_2$ norm of relative residual 
defined as
\[
R_{Global} = || \br ||_2 / || \bF ||_2
\]
The V-cycles are repeated until $R_{Global}  < \ez_G$  where $\ez_G $
is a specified tolerance.

\subsection{Calculation of Elastic Energy Differences}
Our goal is 
given surface site, $(\ell,h_\ell)$, we wish to calculate
the change in the elastic energy 
\[
\Delta W = W ({\bf u}) - W ({\bf u}^a)
\]
where $\bf u$ and ${\bf u}^a$ are the displacement fields with  and 
without the atom, respectively. Table \ref{mg_table} shows the computational
time required to accomplish this task. In more detail we consider a film, with
10 monolayers of deposition, having a profile
similar to the one shown Fig. \ref{fig_test1_mix}.  We begin with an updated
displacement field and remove an atom and compute the change of elastic energy.
The atom is  then replaced; this is done each and every surface atom. Results
for the computational time are displayed on Table \ref{mg_table}. These
were obtained using a 3.6 GHz Intel Pentium 4 Processor Linux Box.
 It should
be pointed out that $ \ez_G =10^{-2}$ provides accurate values
to $\Delta W $ i.e. within 1 to 5 \%.

\begin{table}[h]
\begin{center}
  \begin{tabular}{|l|l|l|l|} \hline
  \multicolumn{4}{|c|}{Fourier-Multigrid CPU Time} \\[.05in] 
  \hline
  System Size       & $ \ez_g =10^{-2}$        & $\ez_g =10^{-3} $ & $\ez_g =10^{-4} $    \\[.05in]
  \hline
   512         &0.030& 0.065&0.202 \\[.02in]
   1024        &0.056&0.128& 0.420\\
  \hline
  \end{tabular}
\end{center}
\caption{Average CPU  time in seconds to update the displacement field and calculate the
change in elastic energy}
\label{mg_table}
\end{table}

\section{Local Elastic Calculations}

Even though the multigrid method greatly reduces the computational time
for an update of the  elastic displacement field it is still far too slow considering
the large number of updates required during the course of a kinetic Monte
Carlo simulation. In this section we will first examine
the long range nature of elastic interactions and then outline a strategy for
the computation of elastic energy differences using local updates
of the displacement field.

\subsection{Long Range Nature of Elastic Interactions}
\begin{figure}[!ht]
\begin{center}
\includegraphics[width=4in]{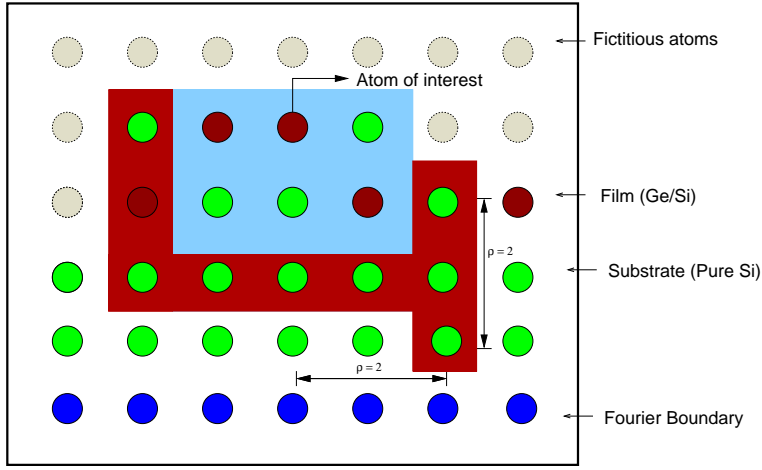}
\end{center}
\caption{\sf \small Computational domain for local
 elastic solve with $\rho=2$ with real and fictitious atoms.
The red circles represent the Germanium atoms and the green circles the Silicon atoms. 
The blue region represents the grid points updated by SOR and
the red region represents the boundary where ${\bf u} = {\bf u_{\rho}}$}
\label{fig:local_domain}
\end{figure}

\begin{figure}[!ht]
\begin{center}
\begin{picture}(300,350)
\put(-50,200){\includegraphics[width=5.5in]{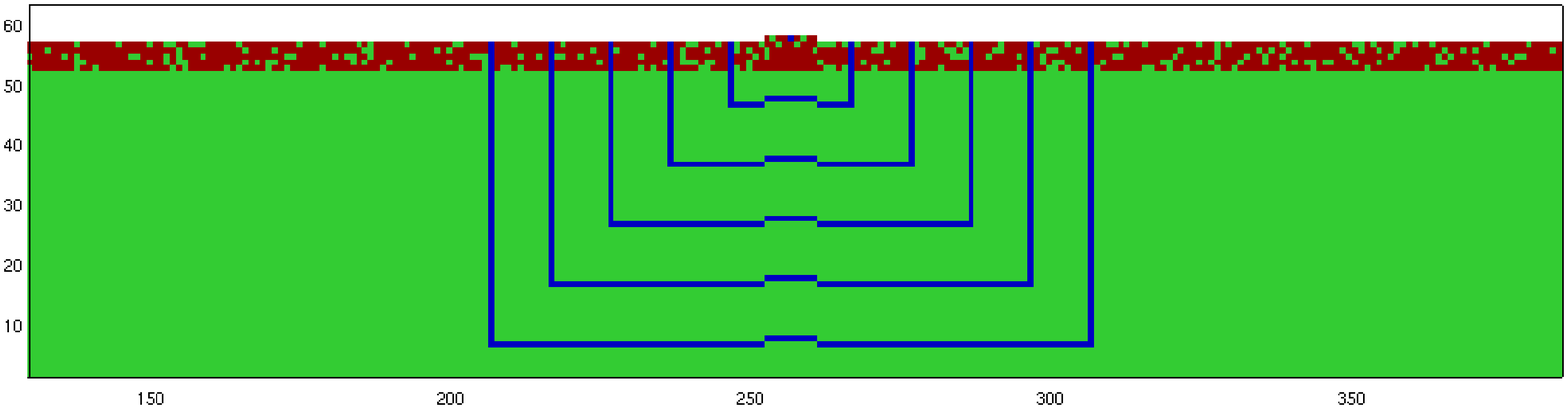}}
\put(-50,0){\includegraphics[width=3in]{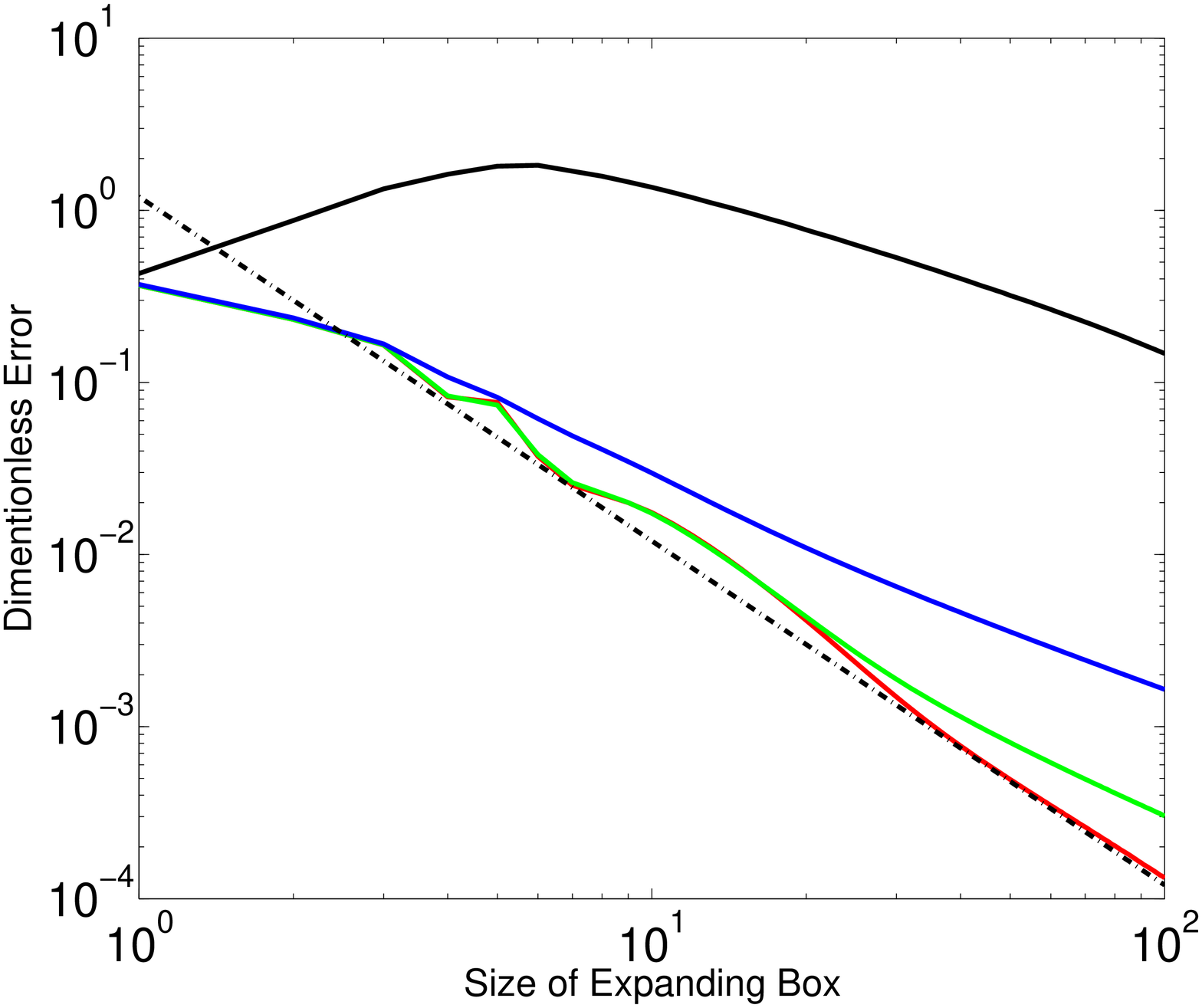}}
\put(153,0){\includegraphics[width=3in]{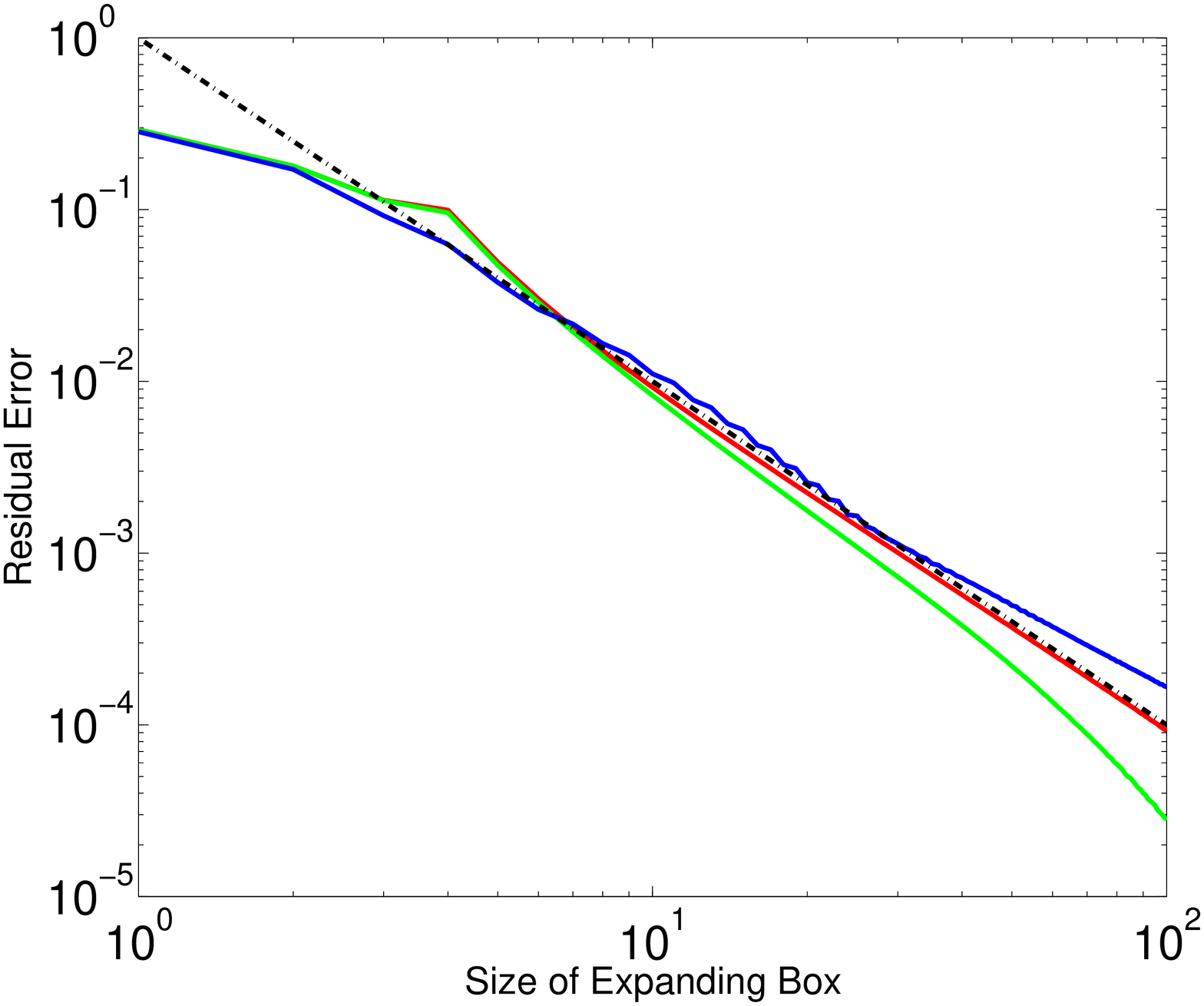}}
\end{picture}
\end{center}
\caption{The upper figure shows a portion of the film along with
some of the expanding boxes. The black line shown on the lower left hand figure
shows a plot of $(\Delta W-\Delta W_{\rho})/ \Delta W $ vs $\rho$.
The colored lines show plots of $(\Delta W-\Delta W_{loc})/ \Delta W $ vs
$\rho$. The blue line is for 2 SOR iterations 
for each box, where green is 10 and red is 50.
The lower right hand figure presents a plot of
$R_{loc}$ for results on the left. In both graphs the dotted line
corresponds to $\rho^{-2}$} \label{fig:case5}
\end{figure}

\begin{figure}[!ht]
\begin{center}
\begin{picture}(300,350)
\put(-50,200){\includegraphics[width=5.5in]{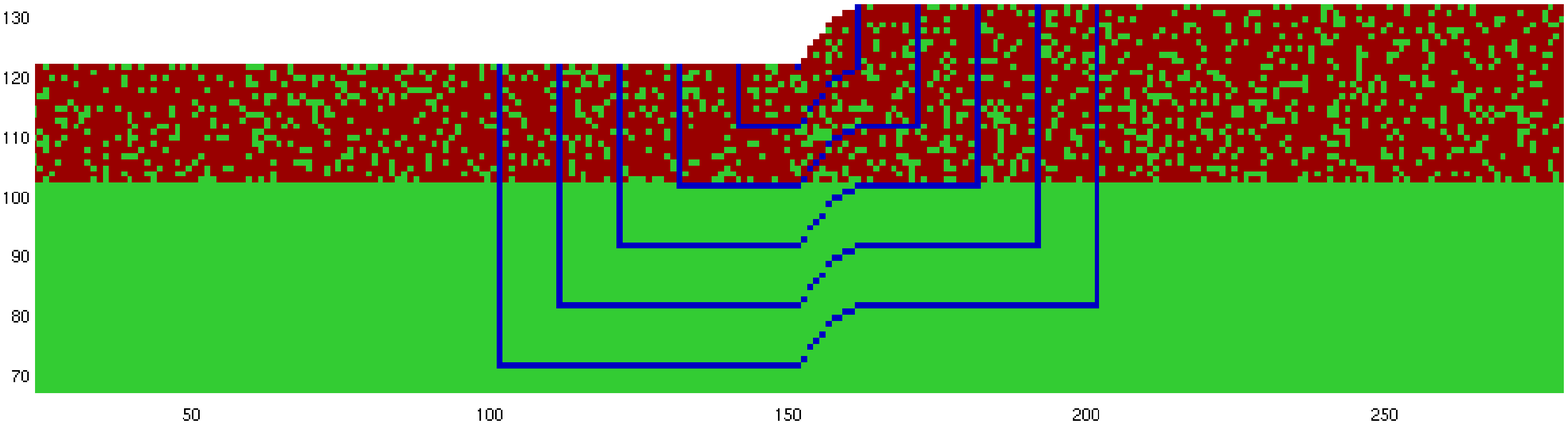}}
\put(-50,0){\includegraphics[width=3in]{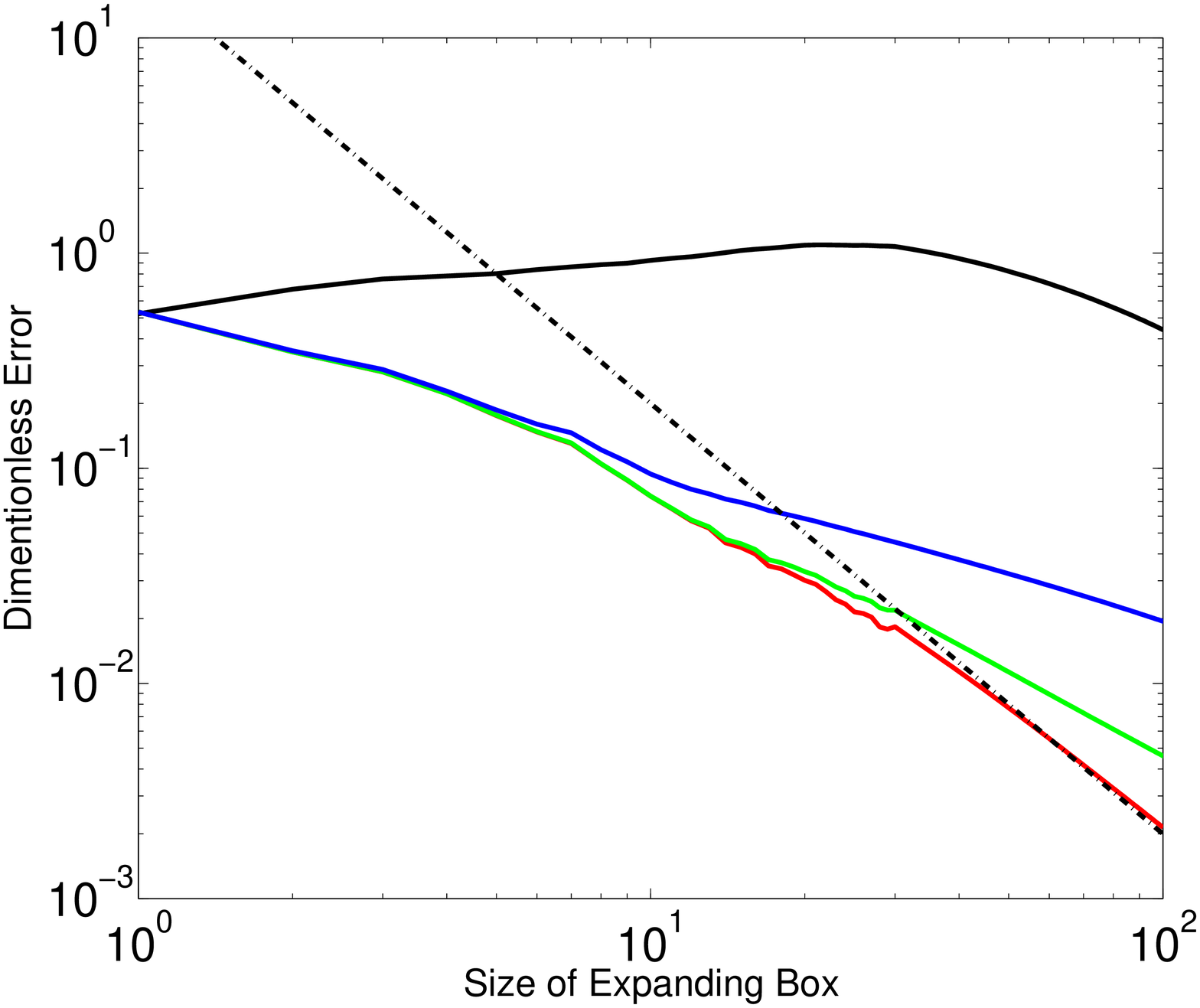}}
\put(153,0){\includegraphics[width=3in]{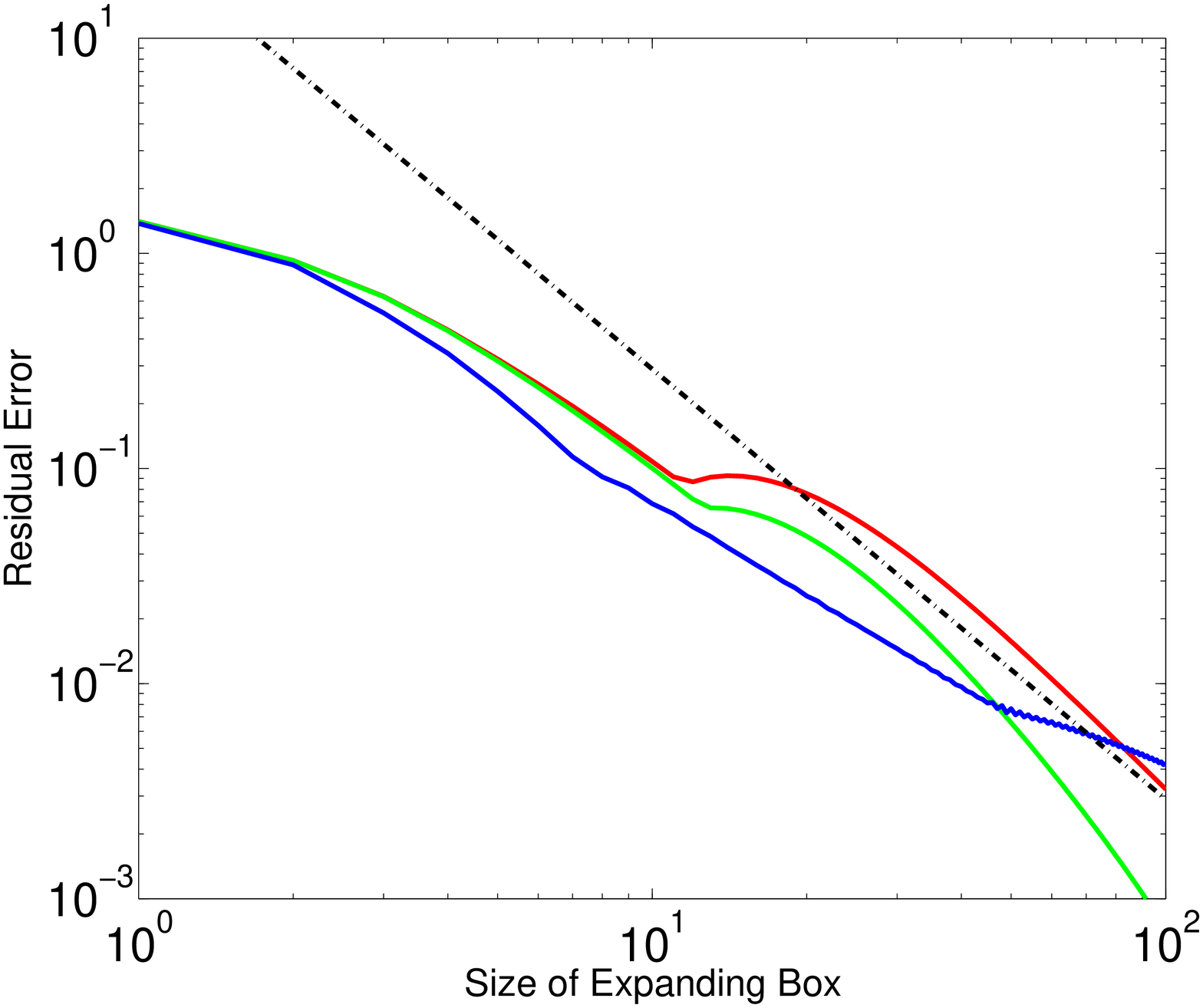}}
\end{picture}
\caption{This is the as Fig. \ref{fig:case5} except
the film profile is different.}
\label{fig:case7}
\end{center}
\end{figure}

As is well known elastic interactions
are long ranged and in fact for strained films they are even longer.
To make this point clear it interesting to consider the following
quantity
\[
\Delta W_\rho = W ({\bf u};\Omega_\rho) - W ({\bf u}^a;\Omega_\rho^a)
\]
where
\[
\Omega_{\rho}:=\{ (\ell,m); i-\rho \leq \ell \leq i+\rho , h_i \geq m \geq h_i-\rho \}
\]
and
\[
\Omega_{\rho}^a:=\Omega_\rho \backslash  \{(i,h_i)\}.
\]
Also in the above formula $W({\bf u};\Omega_{\rho})$ is taken to
mean the total elastic energy inside the region $\Omega\rho$ with
a displacement field given by $\bf u$
as shown in Fig. \ref{fig:local_domain}.
Figures \ref{fig:case5}
and \ref{fig:case7} demonstrate the {\it very} slow
decay of $\Delta W_\rho \to \Delta W$ by plotting
$(\Delta W - \Delta W_\rho)/\Delta W$ vs.
$\rho$ for two different profiles.

These results can be further understood in the context of continuum 
elasticity for a film on semi-finite substrate -- which is a reasonable
approximation since the ball and spring system is a discretization linear elasticity.

We take the film profile to be of the form $h(x)=T+ c(x)$ where
$c(x)$ is a smooth compactly supported function whose
region of support includes $x=0$. If one employs the
small slope approximation and if a small amount of material
is removed from the surface at $(0,h(0))$ then
\begin{equation}
\Delta W_\rho=\Delta W(1+O(T/\rho)) \quad\xtext{as}\quad\rho\to\infty \label{CONT1}
\end{equation}
where $\Omega_\rho$ is a semi-circular region of radius $\rho$ centered at $(0,h(0))$.
For more details see \cite{schulze_smereka}.

\subsection{Principle of Energy Localization}

Given that the goal is to update the displacement field
after removing only one surface atom it might seem reasonable to suppose
that one could locally update the displacement in the vicinity of the
the removed atom. The results above suggest otherwise, however.
Nevertheless, in  \cite{schulze_smereka} the authors present an 
algorithm to perform local updates 
to the elastic fields and calculate the change in the elastic energy, 
with and without the atom, by local calculations.  This
is based on what is called \emph{ Principle of Energy Localization}
\cite{schulze_smereka}  which we will now explain.

The idea behind Energy Localization
is to compute $\Delta W$ by replacing ${\bf u}^a$ by a locally corrected field ${\bf u}^a_{\rho}$
which is computed by solving  Eqs. (\ref{FXC}) and (\ref{FYC})  for atoms
in the domain $\Omega_\rho^a$  using $\bf u$ as Dirichlet boundary data.
In this way we have the following approximate displacement field
for the atom-off configuration
\begin{equation}
{\bf u}^a \approx \left\{ \begin{array}{ll} {\bf u}^a_{\rho} 
& \quad\xtext{for}\quad (\ell,j)\in\Omega_\rho^a\\
                     {\bf u}  & \quad\xtext{otherwise}
                     \end{array}\right.\label{APPROX}
\end{equation}
The principle of energy localization states that
 \[
\Delta W_{loc} = W ({\bf u};\Omega_{\rho}) - W ({\bf u}^a_{\rho};\Omega_{\rho}^a)
\]
will be very close to $\Delta W$ provided that (\ref{APPROX}) is
a good approximation to the displacement field. This result is rather
surprising given that $\Delta W_\rho$ is a terrible approximation and the only
difference between them is that  $ \Delta W_{loc}$ uses a approximation
to ${\bf u}^a$.

To assess the accuracy of (\ref{APPROX}) we note that since both ${\bf u}$ and  ${\bf u}_\rho^a $
satisfy  Eqs. (\ref{FXC}) and (\ref{FYC}) exactly for points not on the boundary of $ \Omega_\rho^a$
then residual will be zero everywhere except for these points.  In other words
the atoms on the boundary of $\Omega_{\rho}^a$ experience a small force.  With this in
mind we define local relative residual  in the region $\Omega_{\rho+1}$  as 
\[
R_{loc} = \frac{1}{\epsilon a_{ss} K_L} \max_{(\ell,j)\in\Omega^a_{\rho+1}} |{\bf r}_{\ell,j}| 
\]
This formula also considers the possibility that the solution inside $ \Omega_\rho^a$
may not satisfy Eqs. (\ref{FXC}) and (\ref{FYC}) exactly.

The following numerical results are consistent with the principle of energy localization
\begin{itemize}
\item $\Delta W_{loc}$ is an excellent approximation to  $\Delta W$ as $\rho$ increases.
\item The local residual $R_{loc}$ decreases as $\rho$ increases.
\end{itemize} 
Figs. \ref{fig:case5} and \ref{fig:case7} show plots of the relative error,
$(\Delta W-\Delta W_{loc})/ \Delta W $ for two different profiles.
The results show that the method can provide accurate values
for the change in elastic energy using local calculations.
 
In addition one can appeal to continuum theory as discussed above to
gain further insight  into the results discuss above. In the same setting
for Eq. (\ref{CONT1}) it is established in \cite{schulze_smereka} that
\begin{equation}
R_{loc} = O(\rho^{-2}) \quad \xtext{as}\quad \rho \to \infty\label{CONT2}
\end{equation}
and that 
\begin{equation}
\Delta W_{loc} = \Delta W (1+ O(\rho^{-2}) ) \quad \xtext{as} \quad \rho \to \infty\label{CONT3}
\end{equation}

\subsection{Expanding Box Method} 

Based on the principle of energy localization Schulze \& Smereka\cite{schulze_smereka} proposed
the Expanding Box Method to construct a local update of the elastic
displacement field that can be used to find an accurate value for the energy
differences. This method is based on using SOR in small neighborhood
of the site where the atom was removed.  When a  change is made to the
crystal configuration at one site, few
iterations of SOR have negligible impact on the solution at sites that
are at a distance of more than one lattice spacing away. 
The expanding box method constructs a series of nested domains 
$\Omega_{\rho}^a$ for $ \rho = 2,3, \ldots ,\rho_{max}$ 
and finds a locally corrected atom-off displacement field, ${\bf u}^a_2$. 
Two iterations of the SOR algorithm described
in \S \ref{SSOR} are used and the atom-on displacement field, $\bf u$,
is used as Dirichlet data for the boundary points of $\Omega_2^a$.
If it is determined that $R_{loc}<\varepsilon_{loc}$, then our
local update of the displacement field has been successful. If not
we repeat the above procedure until $R_{loc}<\varepsilon_{loc}$.
If   $\rho=\rho_{max}$ is reached before a successful local
update has been achieved we resort to a global update using
the multigrid-Fourier method.  

To give some idea on the computation speed of this method we
present some results in Table \ref{EBCPU} for $\rho_{max} =50$. 
Results for $\ez_{loc} =10^{-4}$
were not included since in this case the method fails for almost all attempts.
It is clear that the method is significantly faster than the Fourier-multigrid method.
Two specific examples of this method are
shown on Figs.\ref{fig:case5} and \ref{fig:case7}. The first example shows
an case where the local update would be successful.
In the case shown in Fig. \ref{fig:case7} the decay is slower
than the former case because we are near the base of a large island. In this
example the expanding box method would fail if $\rho_{max} < 50$.
The red curves shown on Figs.\ref{fig:case5} and \ref{fig:case7} are consistent with
the expressions given by Eqs. (\ref{CONT2}) and (\ref{CONT3}) . In addition the
black curves on these figures clearly demonstrate the nonlocal
nature of the elastic energy and is consistent with Eq. (\ref{CONT1}).

In the kinetic Monte Carlo algorithm, we use to simulate strain file growth,
the removed atom is then moved to a neighboring site. Once the atom is moved
the atom-on displacement field, $\bf u$, must be updated. This is also done with the expanding
box method. It should be pointed out that 
the theorems are valid for an initial solution ${\bf u}$ that is exact (in other words a globally computed solution is needed). Nevertheless, in practice this ``quilt'' of locally corrected solutions is found to be
of sufficient accuracy to provide faithful energy differences; this issue
is discussed in more detail in \cite{schulze_smereka}.


\begin{table}
\begin{center}
  \begin{tabular}{|l|l|l|} \hline
  \multicolumn{3}{|c|}{Expanding Box CPU Time} \\ 
  \hline
  System Size       & $ \ez_g =10^{-2}$        & $\ez_g =10^{-3} $ \\
  \hline
   512         &0.0022 (466) & 0.0176 (301)\\
   1024        &0.0022 (930) &0.0106 (608) \\
  \hline
  \end{tabular}
\end{center}
\caption{Average CPU in seconds to update the displacement field and calculate the
change in elastic energy. The number in the parentheses is the number of
successful applications of the expanding box method.}
\label{EBCPU}
\end{table}

\section{Reduced-Rejection Kinetic Monte Carlo}
When using Eqn. \ref{rate_normal_eqn} for parameters of physical interest one finds a wide
range of different rates. In this situation, the most efficient way
to simulate kinetic Monte Carlo is to use a rejection-free implementation.
In this case the hopping rate of every surface atom $R$ needs to be known
before an event can occur.  Even with the expanding box method, this
is too slow for practical simulations. Here we outline a
reduced-rejection approach which reduces the
number of elastic computations.
 
In this method the rates, $R_\ell$, are replaced by
upper bounds, $R^{up}_\ell$ where $R^{up}_\ell\geq R$
where $R^{up}$ can be quickly evaluated.
Now the atoms are selected with rates $R^{up}_\ell$.
Once an atom is selected the actual rate is computed.
To compensate for the overestimate in the rate, the selected atom
will then hop with probability $R_\ell/R^{up}_\ell$. 

This introduces the possibility of rejection whose rate depends on how closely $R_\ell$ 
is approximated by $R^{up}_\ell$. It is clear by looking at (\ref{rate_normal_eqn})
that if we can extract
an appropriate upper bound for the change in elastic energy $\Delta W$
then an upper bound on the rates will follow.  By performing a large
number of numerical experiments, it has been found that $\Delta W$ 
can be bounded above as follows
\[
\Delta W \leq C(N) w_\ell 
\]
where $N$ is the number of occupied neighbors of the $\ell$th surface atom and $w_\ell$
is is the energy stored in the springs attached to this atom.
The function $C(N)$ was found by extensive experiments in \cite{schulze_smereka} 
and was verified for a multi-component film with intermixing 
\[
C(N) = \left\{ 
\begin{array}{ll} 
2.4 &\quad \xtext{if}\quad N=4 \\
3.5 &\quad \xtext{if}\quad N>4 \\
\end{array} \right.
\]
It seems remarkable that $C(N)$ is independent of the type of surface atom
and type of neighboring atoms. The dependence on those factors
is contained in $w_\ell$.

The upper bounds are now, in view of  Eq. (\ref{rate_normal_eqn}), given as
\begin{equation}
R^{up}_\ell = R_o exp\left(\frac{-\gamma N + 
C(N)e_\ell + E_o}{k_B T}\right) \quad \xtext{if} \quad N>3\label{RUP}
\end{equation}
For $N \le 3$ we use the actual rates from Eq. (\ref{rate_normal_eqn}).

\section{The Algorithm}
The KMC algorithm is the same as that given in \cite{schulze_smereka}
but for the convenience of the reader we present it here.

\smallskip
\noindent{\it Precomputation}
\begin{enumerate}
\item
Calculate elastic field of crystal using multigrid-Fourier method.
\item Compute $R_{dep}$ and $R^{up}_\ell$ for $\ell=1$ to $M$ using (\ref{RUP}) 
\end{enumerate}

\bigskip
\noindent{\it Algorithm For Evolution}

\begin{enumerate}

\item Select an event by choosing a uniformly distributed random
number $r \in [0,Z)$, with $Z=R_{dep}+\sum R^{up}_\ell$. This interval
represents an overestimate of the
sum of rates for atoms hopping plus the rate of
deposition.  The event to which $r$ corresponds is located using a
binary tree search \cite{BBS}.

\item If the event is a deposition, locally update the height,
connection arrays and attempt a local elastic solve; revert to a full
elastic solve if the expanding box exceeds size $\rho_{max}$.
Update the rate estimates using (\ref{RUP})
in the same region in which the elastic field was updated.

\item If the event selected is a hop, then
take into account elastic effects by computing the
actual hopping rate $R_\ell \le  R_\ell^{up}$ that depends on the energy difference
of the system with and without the atom.

\begin{enumerate}
\item Make a copy of the atom-on displacement field ${\bf u}$ in
the domain $\Omega_{\rho_{max}}$.
Follow the same procedures in
Step 2 to compute the displacement field with the atom removed,
${\bf u}_{\rho}^a$ (atom off).

\item
Once the elastic field has been updated (locally or globally as necessary),
calculate the energy barrier and actual rate $R_\ell$.

\item
Use rejection to decide whether or not to make the move.
Note that the atom-off
calculation must be performed whether or not this move is made.

\item
If the move is rejected, no change is made to the
displacement field. Return to Step 1.

\item
If the move is
accepted, a hop is made. Update the displacement field  in
the vacated position using ${\bf u}_\rho^a $. Perform a second
local/global calculation in the atom's new position thereby
updating $\bf u$.

\end{enumerate}

\item  Return to Step 1. One event has been completed

\end{enumerate}

\section{Results}
In this section we present some results using the parameters from
\cite{lam_lee_sander} namely, $E_0=0.53 eV$, $R_0 = 2D_0/a_{ss}^2$, 
$D_0 = 3.83 \times 10^{13}\xtext{\AA}^2/sec$, $a_{ss} = 2.73 \xtext{\AA}$,
$\epsilon_{gg}=.04$, $\epsilon_{sg}=.02$.
The spring constants $K_L=13.85 eV/a_s^2$ and $K_D = K_L/2$. 
These were chosen to model the Ge/Si system. In our simulations
we took the bond strength to be $\gamma = 0.37 eV$ whereas
\cite{lam_lee_sander} choose $\gamma =0.4eV$. This was done 
in order to promote the effects of intermixing. It should pointed out
that the above numbers are very approximate and our choice of .37 eV instead of
.4 eV is still well with range of physically reasonable numbers.
The simulations presented here were conducted at a temperature of 600K.
Finally we mention that we take $\ez_G=\ez_{loc}=10^{-2}$.

\begin{figure}[!ht]
\begin{center}
\begin{picture}(300,375)
\put(-80,0){\includegraphics[width=6in]{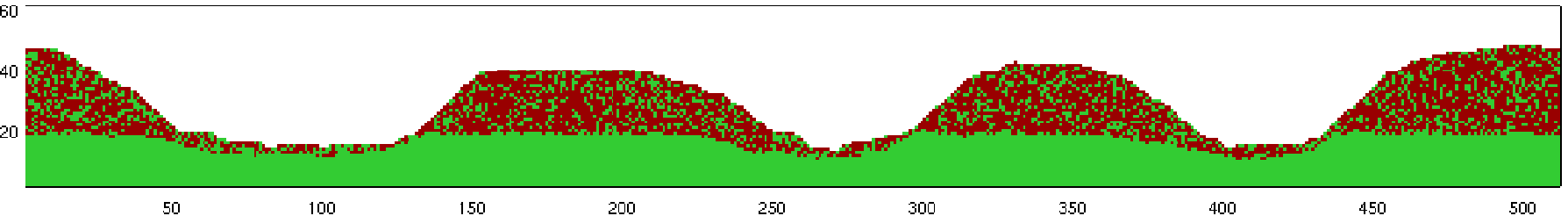}}
\put(-80,75){\includegraphics[width=6in]{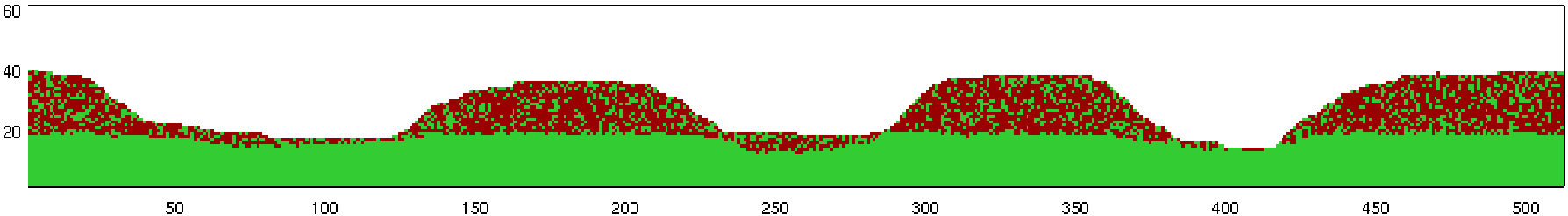}}
\put(-80,150){\includegraphics[width=6in]{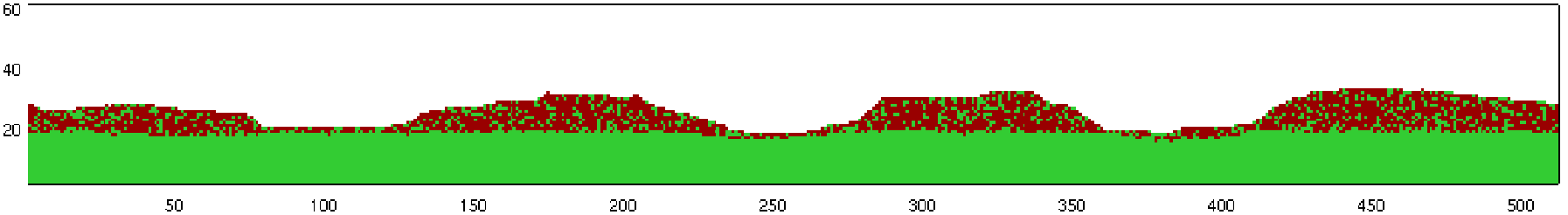}}
\put(-80,225){\includegraphics[width=6in]{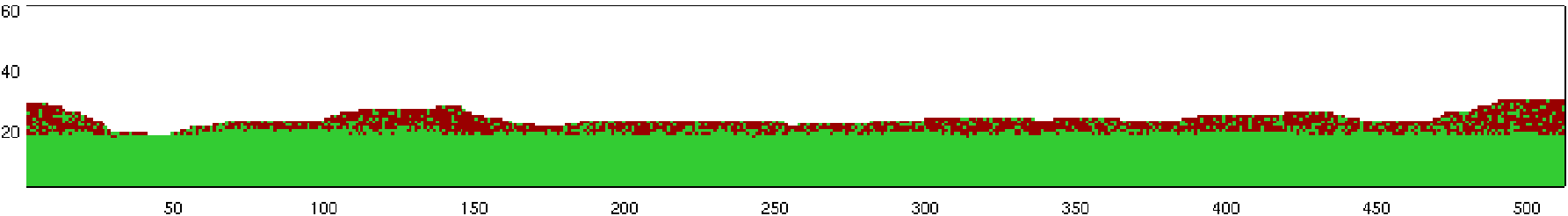}}
\put(-80,300){\includegraphics[width=6in]{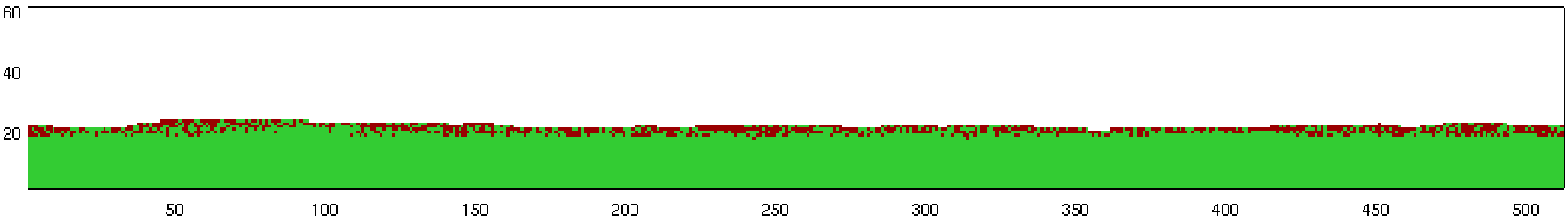}}
\end{picture}
\caption{
Pure Ge on Si : The Ge atoms are represented by red squares and the Si by green squares.
The results show the evolution after 2, 4, 6, 8 and 10 ML deposition.
}\label{fig_test1_mix}
\end{center}
\end{figure}

\begin{figure}[!ht]
\begin{center}
\begin{picture}(300,375)
\put(-80,0){\includegraphics[width=6in]{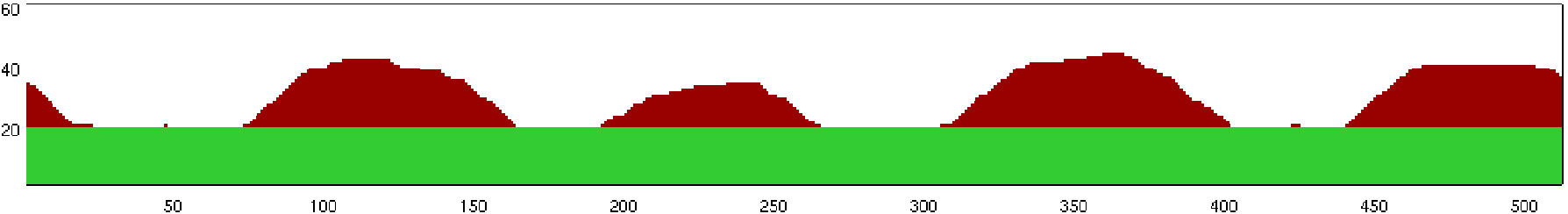}}
\put(-80,75){\includegraphics[width=6in]{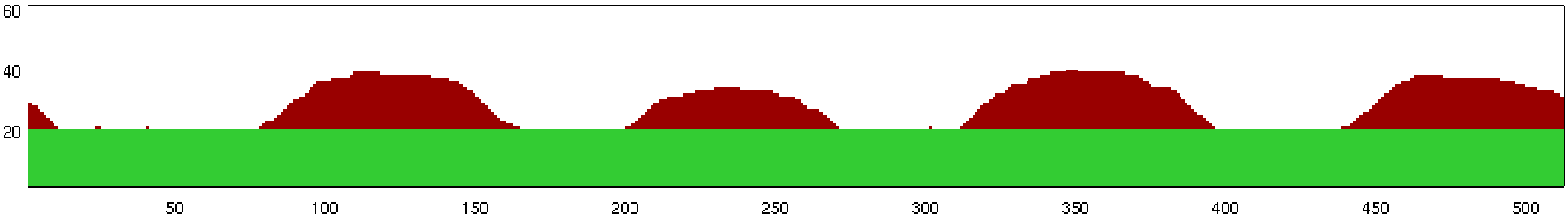}}
\put(-80,150){\includegraphics[width=6in]{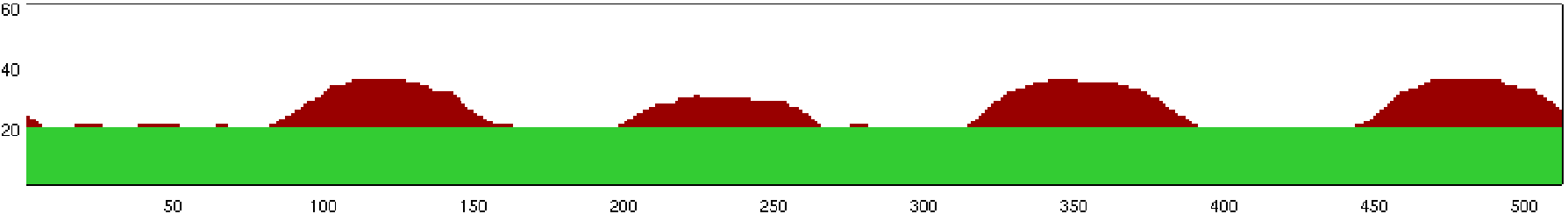}}
\put(-80,225){\includegraphics[width=6in]{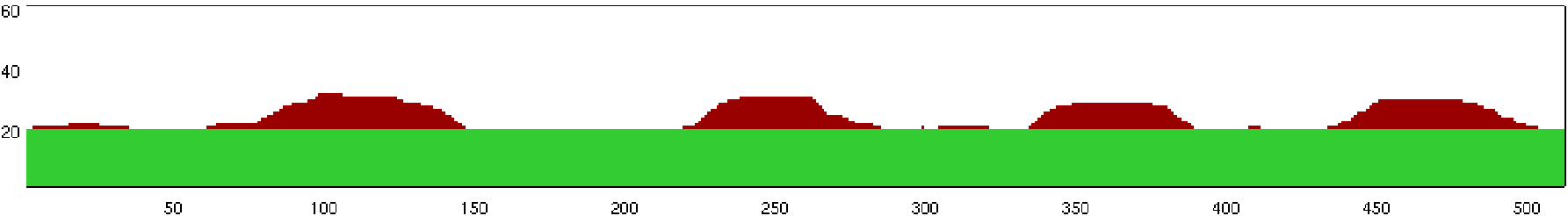}}
\put(-80,300){\includegraphics[width=6in]{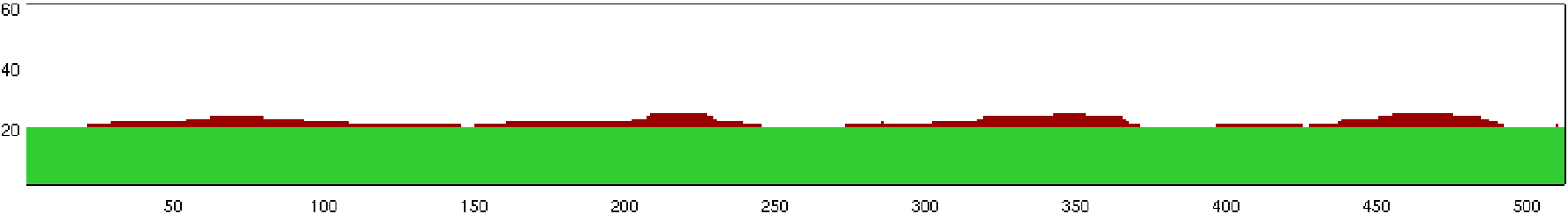}}
\end{picture}
\end{center}
\caption{
Pure Ge on Si : The Ge atoms are represented by red squares and the Si by green squares.
The results show the evolution after 2, 4, 6, 8 and 10 ML deposition.
The Si atoms have been fixed by setting their hopping rates to zero.
}\label{fig_test1_nomix}
\end{figure}

\subsection{Example 1}

The first example outlines the effect of intermixing on
the film evolution. The results are shown in Figs.
\ref{fig_test1_mix} and \ref{fig_test1_nomix}. In both cases
pure Germanium was deposited on pure Silicon substrate
at a rate of $0.8 ML/s$. Fig. \ref{fig_test1_mix} shows the
results of such a simulation. This figure clearly shows the
growth of islands arising from elastic interactions. It also
shows the formation of valleys between islands. The valleys arise
due stress concentration at the base of the islands. The elastic
energy is lowered when the valleys form. During the the
course of the simulation the rejection rate was initially 0.2 \% at 1ML
and increased to 1.14\% at 10ML. In addition, the displacement field
was updated $9.53\times 10^8$ times of which 1679 were global
updates.

Fig. \ref{fig_test1_nomix}
on the other hand shows the same simulation except the
Silicon atoms not allowed to hop. It is clear that the islands form sooner:
in the case with no intermixing there are well defined islands after just
2 monolayers of depositions whereas with intermixing it is closer to 6.
This is because intermixing will
lower the effective misfit thereby weakening the elastic interactions.
A closer inspection reveals that
the pure Ge that was deposited is diluted by around 30 \%.

\begin{figure}[!ht]
\begin{center}
\begin{picture}(200,170)
\put(-150,0){
\includegraphics[width=3in]{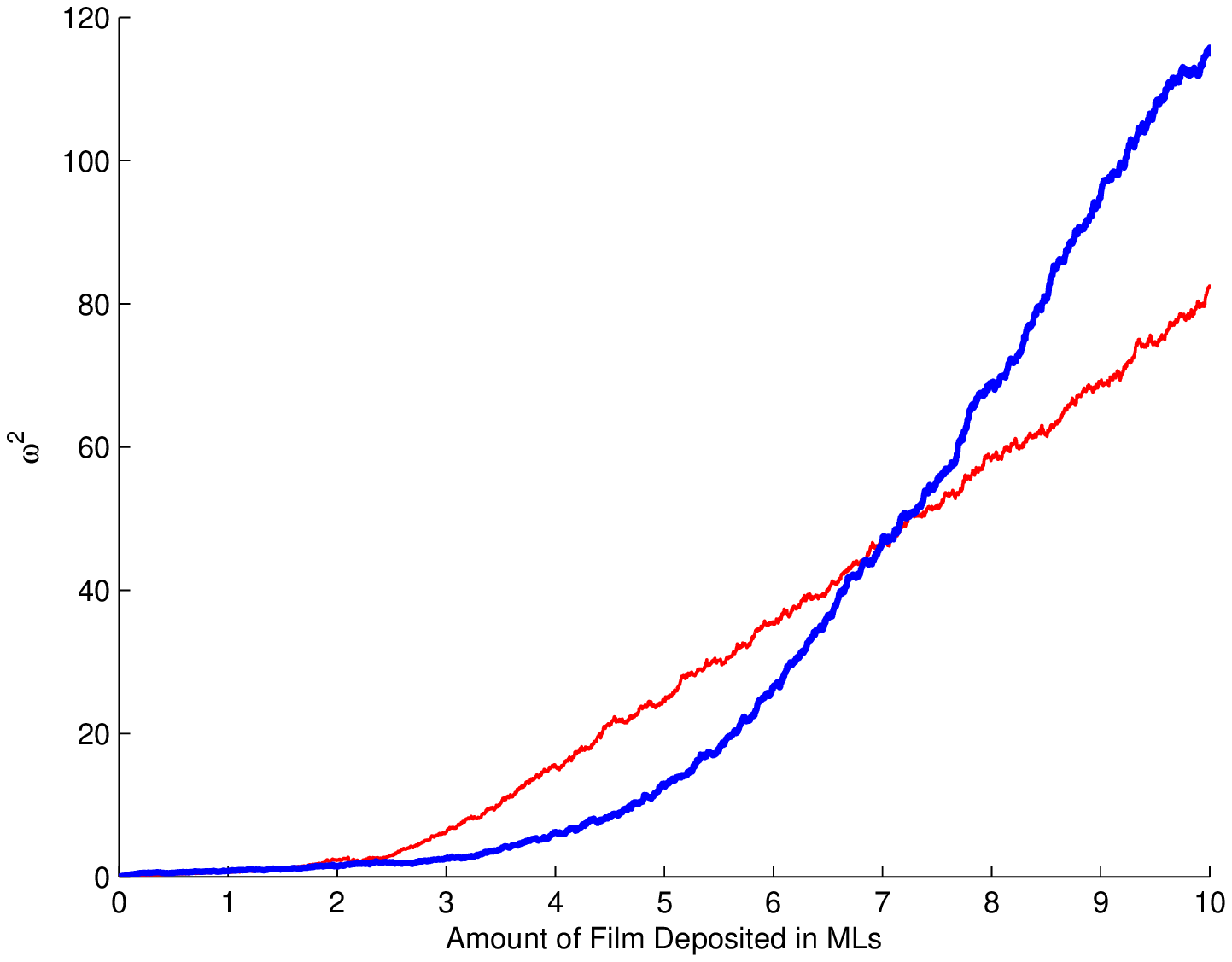}}
\put(100,0){
\includegraphics[width=3in]{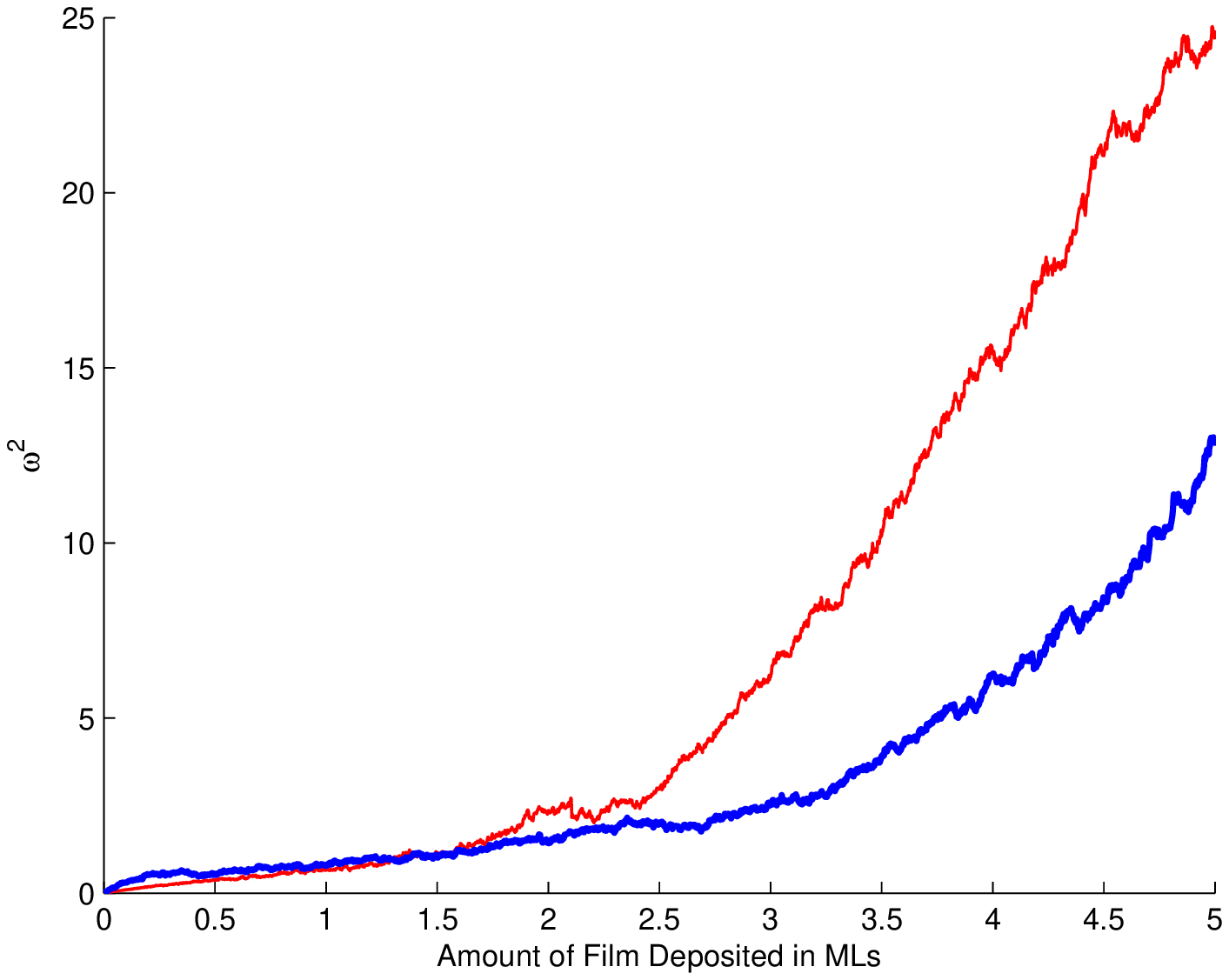}}
\end{picture}
\caption{Roughness Plot : This figure shows the plot of $\omega^2$ the square of the roughness against the film thickness in ML, averaged over 10 independent runs.
The red plot shows the behavior with intermixing and the blue without intermixing. The figure on the right zooms into the first 5 MLs of crystal growth.}\label{fig:roughness}
\end{center}
\end{figure}

To further illustrate the effects of the dilution on the film evolution we present
a plot of the square of the roughness $\omega^2$ (the variance of the height function)
as the function of the film thickness in Fig. \ref{fig:roughness}.
The data represents the ensemble average over 10 different runs. In can be
observed that intermixing initially suppresses growth of the roughness. However
due to the valley formation the roughness in the intermixing case will
ultimately exceed that of the no mixing case.

\noindent

\subsection{Example 2}
\begin{figure}[!h]
\begin{center}
\begin{picture}(300,200)
\put(-70,0){\includegraphics[width=6in]{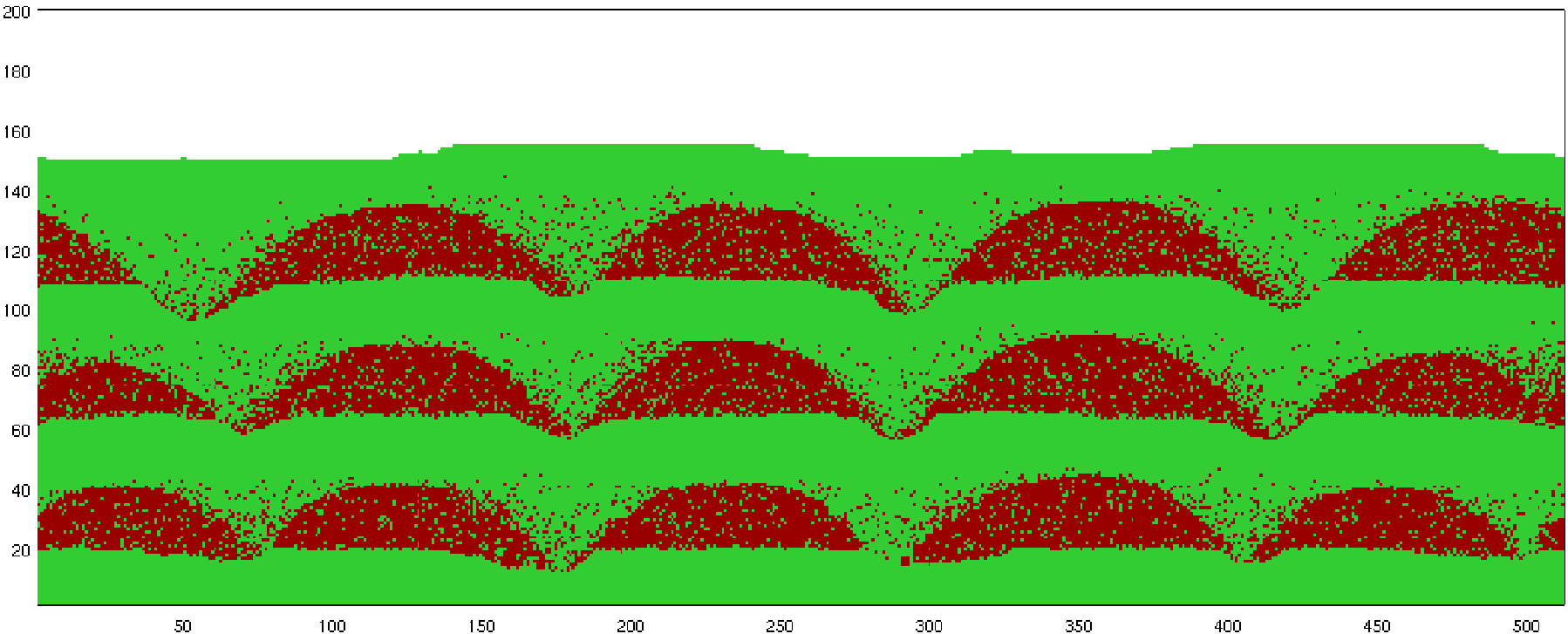}}
\end{picture}
\end{center}
\caption{Stacked Quantum dots: Quantum dots of 15ML Ge with a capping layer of 30 ML Si}
\label{fig:qdots}
\end{figure}
In this example we deposited 15ML of pure Ge on Si at a rate of 10 ML/s
and at a temperature of 600K.  This was capped by 30 ML of Si at the same 
flux and followed by another 15 ML of Ge and so on.
The Ge readily forms islands on the Si substrate. When a new Ge film is grown on the capping 
layer the quantum dots align with the ones buried below.
By this process they self assemble to form an array of stacked 
quantum dots as shown in Fig. \ref{fig:qdots}.

\section{Summary}
In this work, we have presented a kinetic Monte Carlo model for
strained heteroepitaxial growth with intermixing.  The model is based  on 
a solid-on-solid, bond counting formulation \cite{VV} in which elastic interactions are accounted 
for with balls and springs on cubic lattice \cite{orr,lam_lee_sander}. We have also
introduced an algorithm for the efficient  simulation of this model.  This algorithm
is based on the work in \cite{russo_smereka_3D,russo_smereka_2D,schulze_smereka} but
was extended to include intermixing and other improvements. In particular, the
Fourier-multigrid method developed in this work has much neater way of coupling the film
and the substrate as compared to that in \cite{russo_smereka_2D}. In addition,
the course-graining and prolongation operators have been simplified.
We have presented numerical evidence that the principle of
energy localization is valid in the case of intermixing and that the asymptotic results
presented in \cite{schulze_smereka} also remain true. Our
results also indicate that the expanding box method works well in the case
of intermixing. We have also demonstrated that the formula for the upper bounds on the rates
proposed by Schulze \& Smereka\cite{schulze_smereka} appears to work
well even for mixtures. 

A preliminary study on the growth of strained films with intermixing is presented. 
The study indicates the presence of an Asaro-Tiller-Grinfeld type instability of the flat strained film
leading to the formation of quantum dots.
The intermixing is found to considerably lower the strain in the film resulting in a
stabilizing effect that delays the onset of islands. This gives rise
to what Tu \& Tersoff\cite{tersoff_tu} call an apparent critical thickness.
Eventually this critical layer disappears as the valleys form.
A detailed study of the effects of intermixing on the critical thickness is currently under way. 
The model also successfully predicts self assembly of stacked arrays of quantum dots. 

\section*{Acknowledgments}

We thank Len Sander and Tim Schulze for helpful conversations. This work was supported in part by
the following grants from the National Science Foundation, DMS-0553487, DMS-0509124,
and DMS-0810113.

\end{document}